\documentclass[10pt,conference]{IEEEtran}
\IEEEoverridecommandlockouts
\usepackage{cite}
\usepackage{amsmath,amssymb,amsfonts}
\usepackage{algorithmic}
\usepackage{graphicx}
\usepackage{textcomp}
\usepackage[dvipsnames]{xcolor}
\def\BibTeX{{\rm B\kern-.05em{\sc i\kern-.025em b}\kern-.08em
    T\kern-.1667em\lower.7ex\hbox{E}\kern-.125emX}}

\usepackage[linesnumbered,ruled,vlined]{algorithm2e}
\usepackage{hyperref}
\usepackage{tabulary}
\usepackage{tablefootnote}

\usepackage{xurl} %

\usepackage{float}

\usepackage{graphics}
\graphicspath{ {./figs/} }

\usepackage{amsthm}

\usepackage{booktabs}

\usepackage{paralist}

\usepackage{makecell}
\usepackage{multirow}

\usepackage{enumitem}

\usepackage{pgfplots}
\pgfplotsset{compat=1.18}
\usepackage{pgfplotstable}

\usepackage[most]{tcolorbox}
\tcbset{
  colback=green!5!white,
  colframe=green!85!black,
  boxsep=4pt,                 %
  left=0pt, right=0pt,        %
  top=0pt, bottom=0pt,        %
  enhanced,
  sharp corners               %
}

\usepackage{soul}

\newcommand{\step}[1]{%
  \tikz[baseline=(char.base)]{
    \node[shape=circle,fill=black,text=white,inner sep=0pt,minimum size=12pt,align=center] (char) {\sffamily\small #1};
  }%
}

\usepackage{xfp}
\newcommand{\compute}[1]{
    \fpeval{min(#1/28591.0*3,3)}
}

\usepackage{tikz}
\newcommand{\barChart}[2]{%
  \hbox{%
    \begin{tikzpicture}[baseline=(current bounding box.south)]%
      \pgfmathsetmacro{\barwidth}{\compute{#1}}
      \draw[fill=blue!50] (0,0) rectangle (\barwidth,0.18);
    \end{tikzpicture}%
    \hspace{0.2cm}{\footnotesize #2}%
  }%
}

\usetikzlibrary{calc, shapes, arrows.meta, positioning}

\definecolor{color-up-to-date}{HTML}{8CC5E3}
\definecolor{color-tood}{HTML}{2066A8}
\definecolor{color-pfet-tood}{HTML}{F72B8F}

\SetCommentSty{mycommfont}

\newcommand{\metareview}[1]{\textcolor{ForestGreen}{#1}}
\renewcommand{\metareview}[1]{#1}
\newcommand{\review}[1]{\textcolor{ProcessBlue}{#1}}
\renewcommand{\review}[1]{#1}

\newcommand{\goalstatement}{\textit{The goal of this study is to aid developers in making an informed dependency version constraint choice by empirically evaluating the likelihood of dependencies becoming outdated or vulnerable across version constraint types at scale.}\xspace}

\SetKwInput{KwInput}{Input}                %
\SetKwInput{KwOutput}{Output}              %
\SetKwInput{KwNotation}{Notation}

\newcounter{rqcounter}
\newcommand{\newrq}[2]{\noindent\refstepcounter{rqcounter}\textbf{RQ\arabic{rqcounter}:} {\em #2}\label{#1}}
\newcommand{\rqref}[1]{\textbf{RQ\ref{#1}}}

\usepackage{adjustbox}
\usepackage{array}

\newcolumntype{R}[2]{%
    >{\adjustbox{angle=#1,lap=\width-(#2)}\bgroup}%
    l%
    <{\egroup}%
}

\newcommand\pkgname[1]{\textsf{\small #1}}
\newcommand\version[1]{$\operatorname{#1}$}

\newcommand{\pkgdep}{{$<$package, dependency$>$} }
\newcommand{\pkgpkgverdep}{{$<$package, package version, dependency$>$} }

\begin{document}

\title{Which Is Better For Reducing Outdated and Vulnerable Dependencies: Pinning or Floating?
}

\author{

\IEEEauthorblockN{Imranur Rahman, Jill Marley, William Enck, Laurie Williams}
    \IEEEauthorblockA{North Carolina State University
    \\\{irahman3, jahmad5, whenck, lawilli3\}@ncsu.edu}
}

\maketitle

\begin{abstract}

Developers consistently use version constraints to specify acceptable versions of the dependencies for their project.
\emph{Pinning} dependencies can reduce the likelihood of breaking changes, but comes with a cost of manually managing the replacement of outdated and vulnerable dependencies.
On the other hand, \emph{floating} can be used to automatically get bug fixes and security fixes, but comes with the risk of breaking changes.
Security practitioners advocate \emph{pinning} dependencies to prevent against software supply chain attacks, e.g., malicious package updates.
However, since \emph{pinning} is the tightest version constraint, \emph{pinning} is the most likely to result in outdated dependencies.
Nevertheless, how the likelihood of becoming outdated or vulnerable dependencies changes across version constraint types is unknown.
\goalstatement
In this study, we first identify the trends in dependency version constraint usage and the patterns of version constraint type changes made by developers in the npm, PyPI, and Cargo ecosystems.
We then modeled the dependency state transitions using survival analysis and estimated how the likelihood of becoming outdated or vulnerable changes when using \emph{pinning} as opposed to the rest of the version constraint types.
We observe that among outdated and vulnerable dependencies, the most commonly used version constraint type is \emph{floating-minor}, with \emph{pinning} being the next most common.
We also find that \emph{floating-major} is the least likely to result in outdated and \emph{floating-minor} is the least likely to result in vulnerable dependencies.
Based on our findings, we recommend that developers use any kind of \emph{floating} constraint with lockfiles to balance the tradeoffs of \emph{pinning} and \emph{floating}.
\end{abstract}


\section{Introduction}

In open-source software (OSS) ecosystems, developers can specify \emph{version constraints} for a given dependency on which version to depend~\cite{dietrich_dependency_2019}.
\emph{Pinning} and \emph{floating} are two types of specifying \emph{version constraints} for a dependency.
For example, developers can depend \emph{only} on one dependency version (e.g., ``\version{1.2.0}'' of a dependency), which is called \emph{pinning}. Alternatively, they can specify to depend on \emph{version ranges} of dependency (e.g., ``\version{\wedge1.2.0}'' matches \version{1.2.0<=version<2.0.0} of the dependency), which is called \emph{floating}.

In practice, developers have to balance the trade-off between pinning and floating.
With \emph{pinning}, developers can have a deterministic build of their project, which minimizes the likelihood of \emph{breaking changes}~\cite{jafari_dependency_2022} occurring from the dependencies' published versions.
However, developers lose the automatic updates with bug fixes, feature improvements, and security fixes when \emph{pinning} is used.
Suppose a \emph{pinned} dependency is found to be vulnerable according to a security advisory.
In this case, developers have to manually update the vulnerable pinned version to a fixed version of the dependency, ideally after auditing the impact of changes.
Until the fixed dependency version is adopted, the project might be exploited in the wild.
\emph{Pinning} can also lead to packages having outdated dependencies as soon as the upstream developers release a new dependency version.
Having outdated dependencies is risky since they may contain bugs or vulnerabilities~\cite{zerouali_formal_2019}.

On the other hand, \emph{floating} can auto-update the dependency version, including the security fixes.
However, auto-update can introduce breaking changes~\cite{bogart_when_2015,bogart_when_2021} if the upstream developers do not follow semantic versioning (SemVer) appropriately~\cite{raemaekers_semantic_2017}.
For example, developers can choose to \emph{float} the patch version while keeping major and minor versions fixed for a dependency.
However, if the upstream developers remove a method in a patch release, automatically adopting this release may cause breaking changes to the downstream packages~\cite{raemaekers_semantic_2014}.
\emph{Floating} can also have negative security consequences.
For example, an attacker may compromise an OSS package and release a malicious version that is auto-adopted by downstream packages~\cite{ladisa_sok_2023,zahan_openssf_2023,ossf-scorecard}.
In that case, the attacker might be able to exploit the downstream packages.

Developers consistently choose version constraint types while building and releasing their packages' versions, but may not realize the impact of their choices.
Conventional wisdom suggests \emph{pinning} should lead to more outdated dependencies than the rest of the version constraint types since \emph{pinning} is the tightest version constraint type.
However, it remains unknown if \emph{pinning} leads to more vulnerable dependencies in-the-wild.
\metareview{
In addition, He et al.\cite{he_pinning_2025} conducted a study to quantify the cost of maintenance and security benefits using \emph{pinning} and \emph{floating}. We build upon their work with an empirical study on which version constraints are more likely to result in outdated or vulnerable dependencies.
}
Developers do not know \emph{how much more likely} pinning is to result in outdated or vulnerable dependencies compared to the rest of the version constraint types.

\goalstatement
Our in-the-wild analysis can help developers make an informed choice about the version constraint type, considering the security and maintenance implications of their selection.
To reach our goal, we design the following research questions:
\begin{inparaenum}[]
    \item \newrq{rq:versioning}{What is the frequency distribution of different version constraint types in the npm, PyPI, and Cargo ecosystems?}
    \item \newrq{rq:switch}{How frequently do developers change the version constraint type for dependencies, and how does the change result in dependencies being outdated or vulnerable?}
    \item \newrq{rq:impact}{How do pinning and floating affect the dependencies' time to become vulnerable? How do pinning and floating affect the time it takes for dependencies to become outdated?}
\end{inparaenum}

To answer \rqref{rq:versioning}, we use package release data from npm, PyPI, and Cargo, and security advisories to analyze developers' declared version constraints for dependencies.
To answer \rqref{rq:switch}, we consider four transitions in the state of dependencies: (1)~updated to outdated; (2)~outdated to updated; (3)~remediated to vulnerable; and (4)~vulnerable to remediated.
We analyze how frequently developers intentionally change the version constraint type, which then leads to the four transitions.
To answer \rqref{rq:impact}, we model the dependency state transitions using survival analysis to figure out which version constraint type is more likely to result in outdated or vulnerable dependencies.
We use the Cox proportional hazards model to model the \pkgdep relationships and the dependency state transitions to estimate the relative impact of version constraint type on dependency state.

In summary, our study contributes:
(1)~an empirical study of dependency version constraints and their prevalence in npm, PyPI, and Cargo;
(2)~a quantitative analysis of dependency version constraints changes to show trends made by developers in version constraint type;
(3)~a statistical survival analysis to show the relative likelihood of becoming outdated or vulnerable dependencies across version constraint types.
We share our replication package in a Zenodo repository~\cite{zenodo-artifact}.

\section{Background and Definitions}
\label{sec:background}

\subsection{Definitions}
An open source software (OSS) \textbf{package} is a collection of code that provides certain functionality to its users and that is often available in software ecosystem registries (e.g., npm, PyPI) for reuse, modification, and sharing freely~\cite{sonarcube-package}.
When the package is reused by a project or another package, we call the former package a direct \textbf{dependency}.
For example, in Listing~\ref{lst:strategy}, \pkgname{postcss} is the package of interest.
\pkgname{postcss} depends on several other packages, such as \pkgname{commander}, \pkgname{webpack}, and \pkgname{lodash}, and these packages are thus considered dependencies in this context. 
Based on the \textbf{version constraint} specified by the package for the dependencies (e.g., in Listing~\ref{lst:strategy}), the package management software applies a \textit{dependency resolution algorithm} to decide the exact versions of dependencies to be used in the project allowed by \textbf{version constraints}, and we refer to this process as \textbf{dependency resolution}.

When the package uses a version of the dependency that is not the latest available dependency version, we call it an \textbf{outdated dependency}.
When the used dependency version is deemed vulnerable according to a security advisory and a fixed dependency version for the advisory is available, we call it a \textbf{vulnerable dependency}.
\review{
We do not consider unfixed vulnerabilities in our study, since without a fixed version from upstream, there is no easy way for the downstream developers to mitigate the vulnerability.
}

\subsection{Version Constraint Types}
\label{sec:background-constraint-types}

\definecolor{codegreen}{rgb}{0,0.6,0}
\definecolor{codegray}{rgb}{0.5,0.5,0.5}
\definecolor{codepurple}{rgb}{0.58,0,0.82}
\definecolor{backcolour}{rgb}{0.95,0.95,0.92}

\lstdefinestyle{mystyle}{
    backgroundcolor=\color{backcolour},   
    commentstyle=\color{red},
    keywordstyle=\color{magenta},
    numberstyle=\tiny\color{codegray},
    stringstyle=\color{codepurple},
    basicstyle=\ttfamily\footnotesize,
    breakatwhitespace=false,         
    breaklines=true,                 
    captionpos=b,                    
    keepspaces=true,                 
    numbers=left,                    
    numbersep=5pt,                  
    showspaces=false,                
    showstringspaces=false,
    showtabs=false,                  
    tabsize=2,
    string=[s]{"}{"},
    comment=[l]{:\ "},
    morecomment=[l]{\\},
}

\lstset{style=mystyle}

\begin{lstlisting}[label=lst:strategy,float=t,caption={\review{An example package.json file inspired by postcss to illustrate different types of version constraints available in npm, PyPI, and Cargo. More details are present in Section~\ref{sec:background} and in Figure~\ref{fig:constraints-types-taxonomy}.}}]
{ "name": "postcss",
  "version": "0.1.2",
  "dependencies": {
    "commander": "<4.0.0 >1.2.3", \\ fixed-ranging
    "webpack": ">=1.0.0", \\ floating-major
    "meow": "5.x.x", \\ floating-minor
    "jiti": "~3.1.1", \\ floating-patch
    "config": "<9.4.0", \\ at-most
    "dotenv": "5.0.1", \\ pinning
    "postcss-core": "^1.2.0 || ^2.0.1", \\ or-expression
    "cheerio": "!1.2.1", \\ not-expression
    "mocha": ">2.5.0 !2.15 !2.16", \\ complex-expression
    "lodash": "git+ssh://git@github.com:lodash/lodash.git#v.4.11", \\ unclassified
  }, "devDependencies": {...}, ...
}
\end{lstlisting}

\begin{figure}[t]
\centering
\resizebox{0.9\columnwidth}{!}{
\begin{tikzpicture}[
  node distance=0.7cm and 0.7cm,
  every node/.style={font=\LARGE},
  versioning/.style={rectangle, draw, rounded corners=3pt, minimum height=1cm, minimum width=3.2cm, align=center, fill=blue!10},
  other/.style={rectangle, draw, dashed, minimum height=1cm, minimum width=3.2cm, align=center, fill=gray!20}
]

\node[versioning] (pinning) {Pinning};
\node[versioning, right=of pinning] (patch) {Floating Patch};
\node[versioning, right=of patch] (minor) {Floating Minor};
\node[versioning, right=of minor] (major) {Floating Major};

\coordinate (startline) at ($(major.south west) + (-13.4, -1.0)$);
\coordinate (endline) at ($(major.south east) + (0.4, -1.0)$);
\draw[<->, thick] (startline) -- (endline);

\node[anchor=north] at (startline) {\textbf{More Restrictive}};
\node[anchor=north] at (endline) {\textbf{More Flexible}};

\end{tikzpicture}
}
\caption{Spectrum of version constraint types.
}
\label{fig:versioning-spectrum}
\end{figure}
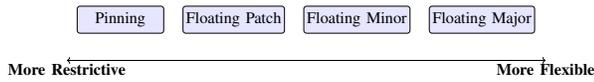

We define a fine-grained categorization of version constraint types below, using the example in Listing~\ref{lst:strategy}.
The relative restrictiveness of the constraint types is illustrated in Fig.~\ref{fig:versioning-spectrum}.

\begin{figure}[t]
    \centering
    \includegraphics[width=\linewidth]{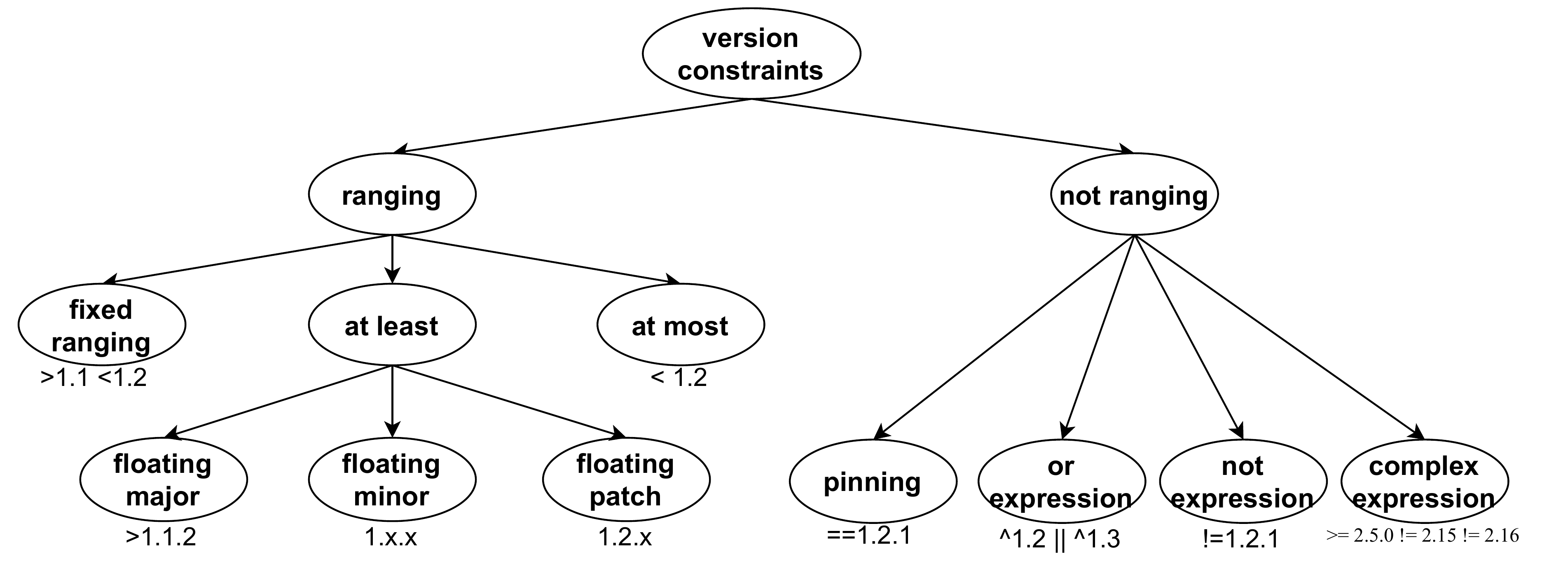}
    \caption{Taxonomy of version constraint types with examples.}
    \label{fig:constraints-types-taxonomy}
\end{figure}

\noindent\textbf{1 Ranging.} If the version constraint contains any kind of version ranges, we call the constraint \emph{ranging}.

\noindent\textbf{1.1 Fixed ranging.} If the version constraint has a fixed version range with upper and lower bounds, we call the version constraint \emph{fixed ranging}.
For example, \pkgname{postcss} specified $<4.0.0 >1.2.3$ for \pkgname{commander} in Listing~\ref{lst:strategy} and this version constraint will resolve to $[>1.2.3, <4.0.0]$.

\noindent\textbf{1.2 At least.} If the version constraint specifies the lower bound and the upper bound is implicit with an operator, we call this version constraint \emph{at least}.
\emph{Floating-major}, \emph{floating-minor}, and \emph{floating-patch} are the three subcategories of \emph{at-least}.

\noindent\textbf{1.2.1 Floating-patch.} If the version constraint of a dependency is specified in a way that the package gets an automatic update when the dependency releases a patch version, we call the constraint \emph{floating-patch}.
For example, \pkgname{postcss} specified $\sim3.1.1$ for \pkgname{jiti} in Listing~\ref{lst:strategy}, and $\sim3.1.1$ will resolve to $[>=3.1.1, <3.2.0]$.

\noindent\textbf{1.2.2 Floating-minor.} If the version constraint of a dependency is specified in such a way that the package gets an automatic update when there is a new minor or patch released by the dependency, we call the constraint \emph{floating-minor}.
For example, \pkgname{postcss} specified $5.x.x$ for \pkgname{meow} in Listing~\ref{lst:strategy}, and $5.x.x$ will resolve to $[>=5.0.0, <6.0.0]$.

\noindent\textbf{1.2.3 Floating-major.} If the version constraint of a dependency is specified in such a way that the package gets an automatic update when there is any new release by the dependency, we call the constraint \emph{floating-major}.
For example, \pkgname{postcss} specified $>=1.0.0$ for \pkgname{webpack} in Listing~\ref{lst:strategy}, and $>=1.0.0$ will resolve to any version greater than \review{or equal to} 1.0.0.

\noindent\textbf{1.3 At most.} If the version constraint specifies the upper bound but no lower bound, we call this version constraint type \emph{at most}.
For example, \pkgname{postcss} specified $<3.4.0$ for \pkgname{config} and $<3.4.0$ will resolve to any version less than $3.4.0$.

\noindent\textbf{2 Not ranging.} If the version constraint does not contain any kind of ranging, we call that \emph{not ranging}.

\noindent\textbf{2.1 Pinning.} If the version constraint of a dependency is fixed to one single version, we call it \emph{pinning}.
For example, \pkgname{postcss} specified $5.0.1$ version constraint for \pkgname{dotenv} in Listing~\ref{lst:strategy}, and the version constraint will resolve to only version $5.0.1$.

\noindent\textbf{2.2 OR expression.} If the version constraint contains a logical OR operator, we call this \emph{or expression}.
In Listing~\ref{lst:strategy}, \pkgname{postcss} specified an \emph{or expression} ``$\wedge1.2.0\ ||\wedge2.0.1$'' for \pkgname{postcss-core}.

\noindent\textbf{2.3 NOT expression.} If the version constraint is specified with the logical NOT operator and without any other operator, we call this \emph{not expression}.
For example, \pkgname{postcss} specified ``$!1.2.1$'' for \pkgname{cheerio} in Listing~\ref{lst:strategy} and the version constraint will resolve to any version of \pkgname{cheerio} except $1.2.1$.

\noindent\textbf{2.4 Complex expression.} If the version constraint contains multiple boolean expressions and multiple operators, or a combination of multiple above types, we call this \emph{complex expression}.
For example, \pkgname{postcss} specified a boolean expression ``$>2.5.0\ !2.15\ !2.16$'' for \pkgname{mocha} in Listing~\ref{lst:strategy}.

\noindent\textbf{2.5 Unclassified.} If the version constraint cannot be placed into any one of the above categories, we call the constraint \emph{unclassified}.
Examples of \emph{unclassified} category could be referring to a version from other repositories (e.g., from GitHub).
For example, \pkgname{postcss} specified $git+ssh://git@github.com:lodash/lodash.git\#v.4.11$ for \pkgname{lodash} in Listing~\ref{lst:strategy}.

\begin{table*}[ht]
    \centering
    \caption{Running example of \pkgname{hexo} package with one of it's depndency \pkgname{moment}.
    }
    \resizebox{\linewidth}{!}{
    \begin{tabular}{|c|c|c|c|c|c|c|c|c|c|c|c|c|}
    \hline
    \makecell[t]{row} & \makecell[t]{pkg} & \makecell[t]{pkg\\version} & \makecell[t]{dep} & \makecell[t]{dep\\constraint} & \makecell[t]{constraint\\type} & \makecell[t]{dep\\version} & \makecell[t]{dep\\highest rel.} & \textbf{\makecell[t]{Interval start}} & \textbf{\makecell[t]{Interval end}} 
    & \textit{\textbf{\makecell[t]{updated}}} & \textit{\textbf{\makecell[t]{remediated}}} 
    \\
    \hline\hline
     & {$\ldots$} & $\ldots$ & $\ldots$ & $\ldots$ & $\ldots$ & $\ldots$ & $\ldots$ & $\ldots$ & $\ldots$ & $\ldots$ & $\ldots$  \\ \hline
    116 & {hexo} & $3.0.1$ & moment & $2.9.0$ & pinning & $2.9.0$ & $2.9.0$ & $\operatorname{2015-04-06}$ & $\operatorname{2015-04-09}$ & true & true \\ \hline
    117 & {hexo} & $3.0.1$ & moment & $2.9.0$ & pinning & $2.9.0$ & $2.10.2$ & $\operatorname{2015-04-09}$ & $\operatorname{2015-05-13}$ & \textcolor{red}{false} & true \\ \hline
    118 & {hexo} & $3.0.1$ & moment & $2.9.0$ & pinning & $2.9.0$ & $2.10.3$ & $\operatorname{2015-05-13}$ & $\operatorname{2015-05-20}$ & \textcolor{red}{false} & true \\ \hline
    119 & {hexo} & $3.1.0$ & moment & $\sim2.10.3$ & floating-patch & $2.10.3$ & $2.10.3$ & $\operatorname{2015-05-20}$ & $\operatorname{2015-05-20}$ & true & true \\ \hline
     & {$\ldots$} & $\ldots$ & $\ldots$ & $\ldots$ & $\ldots$ & $\ldots$ & $\ldots$ & $\ldots$ & $\ldots$ & $\ldots$ & $\ldots$  \\ \hline
    122 & {hexo} & $3.1.1$ & moment & $\sim2.10.3$ & floating-patch & $2.10.6$ & $2.10.6$ & $\operatorname{2015-07-28}$ & $\operatorname{2016-01-02}$ & true & true \\ \hline
    123 & {hexo} & $3.1.1$ & moment & $\sim2.10.3$ & floating-patch & $2.10.6$ & $2.11.0$ & $\operatorname{2016-01-02}$ & $\operatorname{2016-01-09}$ & \textcolor{red}{false} & true \\ \hline
    124 & {hexo} & $3.1.1$ & moment & $\sim2.10.3$ & floating-patch & $2.10.6$ & $2.11.1$ & $\operatorname{2016-01-09}$ & $\operatorname{2016-02-03}$ & \textcolor{red}{false} & true \\ \hline
    125 & {hexo} & $3.1.1$ & moment & $\sim2.10.3$ & floating-patch & $2.10.6$ & $2.11.2$ & $\operatorname{2016-02-03}$ & $\operatorname{2016-02-28}$ & \textcolor{red}{false} & \textcolor{red}{false} \\ \hline
    126 & {hexo} & $3.2.0$ & moment & $\sim2.11.2$ & floating-patch & $2.11.2$ & $2.11.2$ & $\operatorname{2016-02-28}$ & $\operatorname{2016-03-07}$ & true & true \\ \hline
    
    {$\ldots$} & $\ldots$ & $\ldots$ & $\ldots$ & $\ldots$ & $\ldots$ & $\ldots$ & $\ldots$ & $\ldots$ & $\ldots$ & $\ldots$ & $\ldots$ \\
    \end {tabular}
    }
    \label{table:running-example}
\end{table*}

\section{Methodology}
\label{sec:method}

\subsection{Data Collection}
\subsubsection{Package Metadata}
We collected the package dependency relationship information and package version release information from deps.dev~\cite{depsdev} for the npm, PyPI, and Cargo ecosystems on 2024-08-20.
\review{
We chose the three ecosystems for diversity: npm is the largest (5.12M npm vs 715k Maven packages), PyPI is the oldest (PyPI was introduced in 2003 vs Maven Central in 2005), and Cargo is the newest among the major software ecosystems.
}
Package dependency relationship information contains ecosystem, package name, version metadata, e.g., package.json file, and the used direct dependencies along with specified version constraint.
Deps.dev has been used in similar prior studies to collect package metadata in software ecosystems~\cite{shen_understanding_2024,hu_empirical_2024,liu_detecting_2025,alhanahnah_depsrag_2024,akhoundali_morefixes_2024}.
After collecting the data, we had $2,603,314$ npm, $274,720$ PyPI, and $122,069$ Cargo packages before applying any inclusion-exclusion criteria.
To ensure construct quality, we manually inspected a sample of 50 packages, confirming version and dependency metadata against public package registries.
We found deps.dev data to be accurate and consistent with registry data.

\subsubsection{Security Advisories}
We collected security advisories (CVE data) from the CISA Open Source Vulnerabilities (OSV) database, osv.dev~\cite{osv-dev} for npm, PyPI, and Cargo packages on 2024-09-12.
We chose OSV since OSV pulls data from multiple vulnerability feeds, e.g., GitHub Security Advisories~\cite{github-advisory-database}, PyPA, GoVulDB~\cite{osv-data-sources}, for multiple ecosystems and provides the data in a unified OSV format, which facilitates cross-ecosystem analysis.
We have $2,192$ npm, $3,767$ PyPI, and $989$ Cargo security advisories, in total $6,948$.
After collecting the data, we converted it to an SQL table with \emph{advisory identifier}, \emph{ecosystem}, \emph{vulnerable package name}, \emph{version where the vulnerability was introduced}, and \emph{version where the vulnerability was fixed}.
If the vulnerability contains multiple vulnerable version ranges and multiple corresponding fixed versions, we separate the vulnerability into multiple SQL rows where each row corresponds to one SemVer vulnerable version range and one fixed version to facilitate our analysis.

\subsubsection{Package SourceRank}
We examined one characteristic for packages: the SourceRank score, which is available from \review{libraries.io}.
Libraries.io is used by other researchers as a data source~\cite{gu_self-admitted_2023,cao_towards_2023,he_migrationadvisor_2021}.
The Package SourceRank score indicates the package quality, popularity, and community metrics calculated in libraries.io dataset~\cite{saini_investigating_2020,sun_using_2023}.
SourceRank depends on several factors, including the presence of a README file, license, adherence to SemVer, recent updates, and the number of contributors.
We downloaded this data on 2025-01-11 to use as an additional inclusion criterion for ~\rqref{rq:impact}.

\subsubsection{Package Inclusion Criteria}
We begin with an initial dataset of $3,000,103$ ($2,603,314$ npm, $274,720$ PyPI, and $122,069$ Cargo) packages collected from deps.dev.
Our first step is to apply three inclusion criteria: \textbf{\textit{(1)}} the package must be at least two years old, which is operationalized by the time difference between the first and last version release; \textbf{\textit{(2)}} the package must have at least one residual activity (e.g., one version release) in the last two years; and \textbf{\textit{(3)}} the package needs to have at least one dependency.
Our criteria are inspired by Miller et al.~\cite{miller_understanding_2025}'s criteria of abandoned packages since we want to exclude the abandoned packages from our analysis.
Moreover, ``two years'' is a commonly used criterion to measure whether a package is maintained or not~\cite{li_comparison_2023,rahman_no_2025}.
However, our package selection criteria may miss packages that are less than two years old or have had no activity in the last two years (e.g., feature complete packages~\cite{coelho2017modern}).
The number of packages after these inclusion criteria is $163,207$ ($117,129$ npm, $42,777$ PyPI, and $3,301$ Cargo packages).
Among these packages, $22,513$ ($17,263$ npm, $5,158$ PyPI, and $92$ Cargo) packages had at least one vulnerable dependency.

\subsection{Dependency Resolution And Data Preparation}
After collecting the data and applying inclusion criteria, we use the methodology presented by Rahman et al.~\cite{rahman_no_2025} to initially prepare our data for further analysis.
We split each \pkgdep relationship into multiple intervals based on the version release by the package or by the dependency.
By definition, no new version of the package or the dependency is released during each interval.

An example of these constructed intervals (`interval start' and `interval end') is shown in Table~\ref{table:running-example}.
The dependency resolution at each interval takes only the dependency versions available at the beginning of the interval to ensure the dependency resolution accounts for the historical version releases by the dependency.
We use deps.dev~\cite{depsdev} to conduct the dependency resolution with this additional constraint and populate the `dep version' column of Table~\ref{table:running-example}.
\metareview{As identified in Pinckney et al.~\cite{pinckney_large_2023}, npm has a time-travelling feature (\texttt{--before} argument) which can be used to resolve historical dependency resolution.
However, PyPI or Cargo does not have this feature.
To the best of our knowledge, deps.dev is the only service with a dataset that resolves historical dependencies (time-travelling feature) for npm, PyPI, and Cargo.
}
\metareview{Historical dependency resolution for multiple ecosystems is a challenging task, and there is no ground truth for verifying the accuracy of historical dependency resolution by deps.dev.
We manually conducted the historical dependency resolution for 20 packages' versions and their dependencies and created a ground truth.
We then compared our resolution with the resolution provided by deps.dev, and we found that the data from deps.dev matched our ground truth.
}
\review{In addition, dependency constraints with SemVer version qualifiers (e.g., pre-releases, build metadata) are typically not fetched through dependency resolution using version ranges, unless explicitly specified.
So, we exclude these SemVer version qualifiers from our dependency resolution method.
}

We then fill up the boolean columns `updated' and `remediated' to mark if the package uses an outdated or vulnerable dependency by the following conditions.
The resolved version of the dependency is considered \emph{outdated} (`updated'=\textcolor{red}{false}) if the resolved dependency version is not the highest SemVer version of the dependency present at the start of the interval.
Similarly, the resolved version of the dependency is considered \emph{vulnerable} (`remediated'=\textcolor{red}{false}) if the resolved version is within the vulnerable version range for some vulnerability and a fixed version is available at the beginning of the interval.

\subsection{\rqref{rq:versioning}: What is the frequency distribution of different version constraint types in the npm, PyPI, and Cargo ecosystems?}

This RQ consists of two parts: (a) What version constraint types typically occur in general?; and (b) What version constraint types occur in outdated or vulnerable dependencies?

To answer (a), we analyzed the package metadata for all version releases of packages in our dataset.
We classified the used version constraint (`dep constraint') for every package version into one of the version constraint types described in Section~\ref{sec:background-constraint-types} using regular expressions and store that information as another column of the SQL table.
With SQL query, we aggregated the used version constraint type in three granularities: number of unique packages using the type, number of unique \pkgdep using the type, and number of unique \pkgpkgverdep using the type.

To answer (b), which is to understand how dependency-version constraint types relate to outdated or vulnerable dependencies, we analyzed the time intervals during which packages in our dataset depended on outdated or vulnerable versions. 
First, we filtered the time intervals with `updated'=\textcolor{red}{false} on prepared data and calculated another column `time duration' by subtracting `interval start' from `interval end' column.
The result was that for each of the rows, we had the used version constraint type and the associated time duration.
We aggregated the time durations and computed the relative time durations for each version constraint type.
By quantifying the weighted time spent with outdated dependencies across constraint types, we aim to identify which types are most associated with packages' outdatedness.
We followed the same steps with `remediated'=\textcolor{red}{false} for quantifying which types are most associated with packages' vulnerability.

\subsection{\rqref{rq:switch}: How frequently do developers change the version constraint type for dependencies, and how does the change result in dependencies being outdated or vulnerable?}

This RQ explores how changes in version constraints affect whether a dependency becomes outdated or vulnerable.
First (\rqref{rq:switch}a), we want to understand how frequently developers modify the \emph{type of version constraint}.
Then (\rqref{rq:switch}b), we analyze the trends in these version constraint type changes.

\noindent\textbf{\textit{\rqref{rq:switch}a.}}
To understand what \emph{changes} in version constraint and/or version constraint types lead to a package's dependencies being outdated or vulnerable (or vice versa), we analyze four types of state transitions between consecutive time intervals: (1)~updated to outdated; (2)~outdated to updated; (3)~remediated to vulnerable; and (4)~vulnerable to remediated.

\begin{figure}
    \centering
    \includegraphics[width=\linewidth]{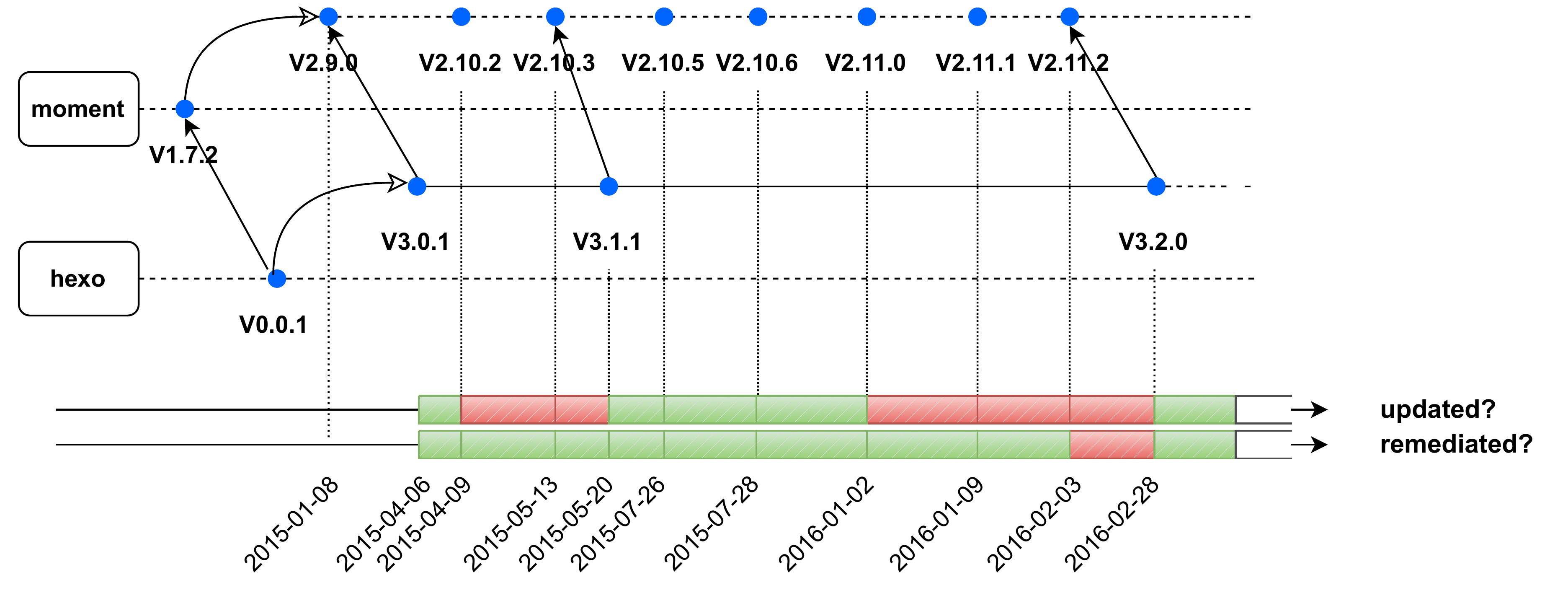}
    \caption{Illustration of dependency state transition using \pkgname{hexo}'s dependency relationship with \pkgname{moment}. \textcolor{green}{Green} time quantum indicates `yes', and \textcolor{red}{red} time quantum indicates `no' in both `updated?' and `remediated?' axis.}
    \label{fig:outdated-vulnerable}
\end{figure}

We illustrate the state transition with the running example in Fig.~\ref{fig:outdated-vulnerable}.
According to our time intervals model, such transitions can occur due to one of the following two events:
\begin{inparaenum}[(i)]
    \item the dependency releases a new version, which the package may or may not adopt, and
    \item the package releases a new version, which may or may not include changes to dependency constraints.
\end{inparaenum}

The above two events can be categorized into four cases based on whether the version constraint type was changed:

\noindent\step{a} The dependency released a new version. For example, if the package depends on a pinned version of a dependency (e.g., $1.2.0$) and the dependency releases a new version $1.2.1$, the package is considered outdated with the release of $1.2.1$.

\noindent\step{b} The package released a new version with the \emph{same dependency version constraint and same constraint type}.

\noindent\step{c} The package released a new version with a \emph{different dependency version constraint but retained the same constraint type} (e.g., remained “pinned” or “floating-minor”).

\noindent\step{d} The package released a new version with a \emph{different dependency version constraint and a changed constraint type} (e.g., switched from “pinned” to “floating-minor”).

Among these four scenarios, \step{a} \review{and \step{b} are the cases} without developers' intervention.
Developers changed the version constraint in \step{c}, and developers changed both the version constraint and version constraint type \step{d}.
So, \step{c} and \step{d} combined represent the cases where developers intervened, and that intervention resulted in a dependency state transition.

\noindent\textbf{\textit{\rqref{rq:switch}b.}}
We further analyze the type \step{d} cases, where the package \emph{changes its dependency constraint type}, since these shifts offer insights into \review{how changes made by developers result in updated or remediated dependencies}.
We examine the prevalence of such \step{d} cases across all four transition types.
For each of the dependency state transitions, we filter out the two consecutive rows for a \pkgdep pair where the package has a different version constraint (`dep version' column) and a different version constraint type (`constraint type' column) in the second row.
We store what changed in such consecutive rows (e.g., \emph{pinned} to \emph{floating-patch}) for the dependency state transition, and aggregate the type change pattern for each of the dependency transitions to compute the prevalence.
These trends reveal how constraint type changes result in dependency freshness and security, directly informing the impact of version constraint types.

\subsection{\rqref{rq:impact}: How do dependencies' version constraint type affect the dependencies' time to become vulnerable? How do dependencies' version constraint type affect the dependencies' time to become outdated?}

\noindent\textit{\textbf{Survival Analysis.}} To answer this RQ, we utilized survival analysis, a statistical methodology exploring time-to-event data.
The expected duration of software projects, or time to termination or delivery of a project, is a popular application of survival analysis in software engineering~\cite{survAnalysisDuration, samoladas2010survival}.
We are interested in observing the time to become vulnerable and the time to become outdated.

\begin{table*}[ht]
\centering
\caption{\metareview{Version Constraint Diversity by Version Constraint Type and Ecosystem }
}
\label{tab:sorted-version-constraint-diversity}
\resizebox{0.95\linewidth}{!}{%
\begin{tabular}{l|rrr|rrr|rrr}
\multirow{2}{*}{\textbf{\makecell[t]{Version Constraint\\Type}}} &
\multicolumn{3}{c|}{\textbf{Cargo}} &
\multicolumn{3}{c|}{\textbf{npm}} &
\multicolumn{3}{c}{\textbf{PyPI}} \\
\cline{2-10}
& \makecell{Unique\\Packages} & \makecell{Unique\\(pkg, dep)} & \makecell{Unique\\(pkg, pkgver, dep)} 
& \makecell{Unique\\Packages} & \makecell{Unique\\(pkg, dep)} & \makecell{Unique\\(pkg, pkgver, dep)} 
& \makecell{Unique\\Packages} & \makecell{Unique\\(pkg, dep)} & \makecell{Unique\\(pkg, pkgver, dep)} \\
\midrule

floating-minor     & \textbf{3,268 (89.12\%)} & \textbf{6,674 (92.22\%)} & \textbf{99,877 (95.29\%)}  & \textbf{105,826 (62.63\%)} & \textbf{614,392 (71.86\%)} & \textbf{34,092,953 (66.72\%)} & 249 (0.47\%) & 487 (0.25\%) & 5,749 (0.13\%) \\
pinning            & 141 (3.85\%)  & 192 (2.65\%) & 2,986 (2.85\%)  & 41,216 (24.39\%)  & 180,851 (21.15\%) & 15,384,917 (30.11\%) & 9,016 (17.06\%)  & 47,036 (24.33\%) & 1,286,540 (29.40\%) \\
floating-patch     & 111 (3.03\%)  & 173 (2.39\%)  & 1,302 (1.24\%)   & 9,533 (5.64\%)   & 28,500 (3.33\%)  & 1,100,620 (2.15\%)  & 3,375 (6.38\%)  & 11,551 (5.98\%) & 237,391 (5.43\%) \\
fixed-ranging      & 125 (3.41\%)  & 176 (2.43\%)  & 535 (0.51\%)  & 795 (0.47\%)   & 1,383 (0.16\%)   & 75,018 (0.15\%)   & 12,823 (24.26\%) & 48,638 (25.16\%) & 1,265,353 (28.92\%) \\
at-most            & 1 (0.03\%)    & 1 (0.01\%)    & 1 (0.00\%)       & 136 (0.08\%)    & 187 (0.02\%)    & 3534 (0.01\%)     & 3,291 (6.23\%)  & 5,523 (2.86\%)  & 74,123 (1.69\%) \\
complex-expression & 3 (0.08\%)    & 3 (0.04\%)    & 3 (0.00\%)       & 19 (0.01\%)     & 29 (0.00\%)     & 626 (0.00\%)      & 1,773 (3.35\%)  & 3,551 (1.84\%)  & 49,113 (1.12\%) \\
or-expression      & 0 (0.00\%)    & 0 (0.00\%)    & 0 (0.00\%)       & 1,225 (0.73\%)   & 1,859 (0.22\%)   & 40,702 (0.08\%)    & 0 (0.00\%)     & 0 (0.00\%)     & 0 (0.00\%) \\
floating-major     & 18 (0.49\%)  & 18 (0.25\%)  & 113 (0.11\%)  & 10,220 (6.05\%)  & 18,757 (2.19\%)  & 402,644 (0.79\%)  & \textbf{22,087 (41.78\%)} & \textbf{76,232 (39.43\%)} & \textbf{1,453,788 (33.23\%)} \\
not-expression     & 0 (0.00\%)    & 0 (0.00\%)    & 0 (0.00\%)       & 0 (0.00\%)      & 0 (0.00\%)      & 0 (0.00\%)        & 246 (0.47\%)   & 296 (0.15\%)   & 3,449 (0.08\%) \\
\midrule
total  &  3,667  & 7,237  & 104,817  & 168,970  & 854,958  & 51,101,014  & 52,860  & 193,314  & 4,375,506 \\
\end{tabular}
}
\end{table*}

\noindent\textit{\textbf{Model Selection.}}
The Cox proportional hazards model is a survival method utilized to examine the relationship between the time until an event occurs and a set of explanatory variables \cite{dawson2021application}.
The use of this model enables researchers to estimate the relative risk between groups, which is expressed in hazard ratios.
The hazard function, also known as the hazard rate, is the conditional probability of an individual experiencing the event of interest, given the event has not yet occurred up to that point in time \cite{bewick2004statistics}.
The hazard ratio is the ratio of the hazard rate for the treatment group to that of the baseline group \cite{steyerberg2010concepts}. 

In our study, we are interested in evaluating the relationship between dependencies' version constraint types and the hazards of becoming outdated or vulnerable.
We constructed two models: one with time to vulnerability as the event and one with time to outdated as the event.
The covariates in our models were the version constraint types.
We chose \emph{pinning} as the baseline group as \emph{pinning} is the most restrictive type, allowing for a straightforward interpretation of the relative hazard ratios.
Because several dependencies transitioned between constraint types (for example, 85 dependencies transition from \emph{floating-major} to \emph{pinning}), we used the time-varying Cox proportional hazards model (from Python's lifelines~\cite{Davidson-Pilon2019} package), which enables covariates to change their values over time \cite{zhang2018time}.

\noindent\textit{\textbf{Dataset construction}}
We filtered our dataset for the top 1000 packages of each ecosystem (a total of 3000 packages across the three ecosystems) based on SourceRank scores obtained from the libraries.io dataset.
After filtering, our initial survival dataset consisted of 16,223 unique dependencies from the $<$ecosystem, package, dependency$>$ relationship, with over 1.5M rows of data.
Each row of data consisted of the dependency ID, an interval start time, an interval end time, and the version constraint type.
We removed any dependency where the interval starting time was the same as the interval ending time, as these rows may cause a spurious result.
Our final survival dataset consisted of 16,094 unique dependencies and \review{contained data points for both changed constraints and unchanged constraints.}
The dependent variables in our models were whether or not the dependency was outdated and whether or not the dependency was vulnerable. %

\section{Results}
\label{sec:results}

\subsection{\rqref{rq:versioning}: What is the frequency distribution of different version constraint types in the npm, PyPI, and Cargo ecosystems?}

To answer this RQ, we analyze the version constraint types used across npm, PyPI and Cargo.
We divide our analysis into two parts:
\textit{(a) What version constraint types typically occur in general?} 
Table~\ref{tab:sorted-version-constraint-diversity} shows how frequently different version constraint types appear in each ecosystem.
We count the use of each version constraint type in three granularities: the number of unique packages, unique $<$package, dependency$>$ pairs, and unique $<$package, version, dependency$>$ tuples.
We can see that \emph{floating-minor} is the most commonly used in npm and Cargo.
\emph{Floating-minor} is used over 99k times in $<$package, version, dependency$>$ granularity in Cargo, 34M times in npm, and 6k times in PyPI.
\emph{Pinning} is the second most frequently used type in Cargo (3k times), npm (15M times), and PyPI (1.2M times).
However, in PyPI, \emph{floating-major} is the most commonly used, followed by \emph{pinning} and \emph{fixed-ranging}, which differ from npm and Cargo.
\emph{Floating-patch}, \emph{at-most}, and \emph{complex-expression} appear less frequently.
Ecosystem developers have certain version constraint types that are unique to that ecosystem.
For example, \emph{or-expression} occurs only in npm and \emph{not-expression} occurs only in PyPI.

\noindent \textit{(b) What version constraint types typically occur in outdated and in vulnerable dependencies?}
We computed the prevalence of constraint types associated with outdated dependency versions and vulnerable dependency versions.
Table~\ref{tab:occurence-percentage} shows the distribution of version constraint types among outdated and among vulnerable dependencies.
We can see that \emph{floating-minor} and \emph{pinning} account for the majority of time intervals with outdated and with vulnerable dependencies.
\metareview{
\emph{Floating-minor} represents 95\% of the dependencies in Cargo but accounts for 95\% outdated and 99\% vulnerable dependencies (\textit{over-represented}).
Also, \emph{floating-minor} represents 67\% of the dependencies in npm, but accounts for 67\% outdated and 56\% vulnerable dependencies (\textit{under-represented}) in npm.
\emph{Pinning} represents 30\% of the dependencies in npm but accounts for 28\% outdated (\textit{under-represented}) and 38\% vulnerable dependencies (\textit{over-represented}).
In addition, \emph{pinning} represents 29\% of the dependencies in PyPI but accounts for 56\% outdated (\textit{over-represented}) and 66\% vulnerable dependencies (\textit{over-represented}).
Moreover, \emph{fixed-ranging} represents 29\% dependencies in PyPI but accounts for 30\% outdated (\textit{over-represented}) and 23\% vulnerable dependencies (\textit{under-represented}).
However, \emph{fixed-ranging} accounts for a small portion of outdatedness and vulnerability in Cargo and npm, given that \emph{fixed-ranging} constitutes 0.51\% of Cargo and 0.14\% of npm dependencies.
Over-representation indicates the constraint is more likely to result in outdated or vulnerable dependencies in that ecosystem.
Similarly, under-representation indicates the constraint is less likely to result in outdated and vulnerable dependencies in that ecosystem.
}

\noindent The rest of the constraint types contribute much less to outdatedness or vulnerability, each with less than 6\%.
Less frequent version constraint types (from \rqref{rq:versioning}a), e.g., \emph{or expression}, \emph{not expression}, and \emph{complex expression}, are less frequent in outdated and vulnerable dependencies as well.

\begin{table}[t]
\centering
\caption{Occurrence Percentage of Version Constraint Types in Outdated and in Vulnerable Dependencies.}
\label{tab:occurence-percentage}
\resizebox{0.95\linewidth}{!}{
\begin{tabular}[t]{lrrrrrr}
\toprule
\textbf{\makecell[t]{Version Constraint \\Type}} & \multicolumn{2}{c}{\textbf{Cargo}} & \multicolumn{2}{c}{\textbf{npm}} & \multicolumn{2}{c}{\textbf{PyPI}} \\
\cmidrule(lr){2-3} \cmidrule(lr){4-5} \cmidrule(lr){6-7}
 & \makecell{Outdated\\($\%$)} & \makecell{Vulnerable\\($\%$)} & \makecell{Outdated\\($\%$)} & \makecell{Vulnerable\\($\%$)} & \makecell{Outdated\\($\%$)} & \makecell{Vulnerable\\($\%$)} \\
\midrule
floating-minor     & \textbf{95.55} & \textbf{99.09} & \textbf{67.74} & \textbf{56.20} & 0.16 & 0.07 \\
pinning            & 2.64  & 0.04  & 27.79 & 37.73 & \textbf{56.02} & \textbf{65.83} \\
floating-patch     & 1.33  & 0.81  & 4.27  & 5.93  & 8.14 & 7.37 \\
fixed-ranging      & 0.45  & 0.06  & 0.08  & 0.07  & 29.57 & 22.73 \\
at-most            & 0.02  & -     & 0.02  & 0.05  & 4.56 & 3.55 \\
complex-expression & 0.00  & -     & 0.00  & 0.00  & 0.59 & 0.45 \\
or-expression      & -     & -     & 0.10  & 0.03  & -    & -    \\
floating-major     & -     & -     & -     & 0.00  & 0.95 & 0.00 \\
not-expression     & -     & -     & -     & -     & 0.02 & 0.00 \\
\bottomrule
\end{tabular}
}
\end{table}

\noindent\textbf{Why is \emph{floating-minor} the most used in outdated dependency?}
\metareview{
First of all, \emph{floating-minor} is the most frequently used dependency constraint type.
Since \emph{floating-minor} is the most represented in the dependency constraints, \emph{floating-minor} is expected to be the most represented in both outdated and vulnerable dependencies.
}
Second, based on our definition of outdated dependency, we compared the used dependency version in each interval with the highest available version of the dependency, which is considered the ``ideal'' dependency version in our case.
This choice of ``ideal'' dependency version leads to \emph{floating-minor} being the most used in outdated dependencies.
If a package uses \emph{floating-minor} for a dependency and the dependency releases a new major version, the package will continue to have ``outdated'' dependency until the package releases a new version adopting the newest major version of the dependency.
Similarly, if the upstream developers maintain multiple major branches, and the downstream developers use \emph{floating-minor} with an older major branch, the package will always be considered ``outdated'' for that dependency.

\begin{table}[t]
    \centering
    \caption{Prevalence of constraint type changes leading to dependency state transitions.}
    \resizebox{\linewidth}{!}{
    \begin{tabular}{|l|r|r|r|r|}
    \hline
    \textbf{Transition Type} & \textbf{\step{a}} & \textbf{\step{b}} & \textbf{\step{c}} & \textbf{\step{d}} \\
    \hline
    1. Updated \ \ \ \ $\rightarrow$ Outdated & 246 & 1,016 & 47,795 & 22,887 \\ \hline
    2. Outdated \ \ \ $\rightarrow$ Updated & 5,645 & 85 & 5,222,276 & 67,681 \\ \hline
    3. Remediated $\rightarrow$ Vulnerable & 16 & 1,271 & 594 & 1,133 \\ \hline
    4. Vulnerable \ $\rightarrow$ Remediated & 51 & 0 & 19,549 & 3,829 \\ \hline
    Total count                          & 5,958 & 2,372  & 5,290,214 & 95,530 \\ \hline
    Column \%                      & 0.11\% & 0.04\% & 98.09\% & 1.77\% \\
    \hline
    \end{tabular}
    }
    \label{table:strategy-prevalence}
\end{table}

\begin{table*}[ht]
\centering
\caption{Top 5 Most Frequent Version Constraint Changes by Dependency Transition Type}
\begin{tabular}{|l|l|l|c|l|}
\hline
\textbf{Dependency Transition} & \textbf{From Constraint} & \textbf{To Constraint} & \textbf{Constraint Change} & \textbf{Frequency} \\
\hline

\multirow{5}{*}{Updated \ \ \ \ $\rightarrow$ Outdated} 
& floating-minor & pinning & $\uparrow$ constrained & \barChart{12853}12,853 \\
& floating-major & fixed-ranging & & \barChart{2739}2,739 \\
& floating-major & pinning & $\uparrow$ constrained & \barChart{2243}2,243 \\
& floating-minor & floating-patch & $\uparrow$ constrained & \barChart{1630}1,630 \\
& fixed-ranging & pinning & & \barChart{660}660 \\
\hline

\multirow{5}{*}{Outdated \ \ \ $\rightarrow$ Updated} 
& pinning & floating-minor & $\downarrow$ constrained & \barChart{28591}28,591 \\
& floating-minor & pinning & $\uparrow$ constrained & \barChart{9424}9,424 \\
& pinning & floating-major & $\downarrow$ constrained & \barChart{5448} 5,448 \\
& floating-patch & floating-minor & $\downarrow$ constrained & \barChart{4442} 4,442 \\
& fixed-ranging & floating-major &  & \barChart{4006}4,006 \\
\hline

\multirow{5}{*}{Remediated $\rightarrow$ Vulnerable} 
& floating-minor & pinning & $\uparrow$ constrained &\barChart{541} 541 \\
& floating-major & pinning & $\uparrow$ constrained & \barChart{199}199 \\
& floating-major & fixed-ranging & &\barChart{117} 117 \\
& fixed-ranging & pinning & $\uparrow$ constrained &\barChart{61} 61 \\
& floating-minor & floating-patch & $\uparrow$ constrained &\barChart{43} 43 \\
\hline

\multirow{5}{*}{Vulnerable \ $\rightarrow$ Remediated} 
& pinning & floating-minor & $\downarrow$ constrained & \barChart{1154}1,154 \\
& pinning & floating-major & $\downarrow$ constrained & \barChart{694}694 \\
& pinning & fixed-ranging & $\downarrow$ constrained & \barChart{342}342 \\
& fixed-ranging & floating-major & & \barChart{331}331 \\
& floating-minor & pinning & $\uparrow$ constrained & \barChart{278}279 \\
\hline

\end{tabular}
\label{tab:spec-strategy-transitions}
\end{table*}

\noindent\textbf{Why \emph{floating-minor} is the most used in vulnerable dependency?}
For each row of our dataset, we examine if the used dependency version is vulnerable and whether a fixed version is available at the start of the interval; if yes, we consider the used dependency version vulnerable.
Based on SemVer policy~\cite{semver-spec,pinckney_large_2023,li_large-scale_2023}, we do not expect \emph{floating-minor} to be the most frequently used version constraint type in vulnerable dependencies.
So we hypothesize that packages get security fixes automatically when they use \emph{floating-minor}.
To verify our hypothesis,
we analyzed the vulnerable version ranges (e.g., SemVer versions from vuln introduced to vuln fixed) associated with vulnerabilities in our dataset.
We manually reviewed 50 randomly chosen CVEs to validate our hypothesis.
We found that, as opposed to our hypothesis, \emph{vulnerability fixes are not being propagated to all major vulnerable versions}.

In our dataset, we found 4,219 (2,465 npm, 1,560 PyPI, and 194 Cargo) CVEs and 1,553 (473 npm, 1,001 PyPI, and 79 Cargo) packages where the vulnerable versions are spread across multiple major versions, but only one major version received the fix.
As a case study, we present \href{https://osv.dev/vulnerability/GHSA-23wx-cgxq-vpwx}{GHSA-23wx-cgxq-vpwx} where package \pkgname{dset}'s version [0.0.0 - 3.1.1] are vulnerable, and the fixed version for this vulnerability is 3.1.2.
Although the package \pkgname{dset} has versions \{0.0.0, 1.0.0, 1.0.1, 2.0.0, 2.0.1, 2.1.0, 3.0.0, 3.1.0, 3.1.1, 3.1.2, ...\}, the fix is not propagated to the major versions \version{0}, \version{1} and \version{2}.
Not propagating the vulnerability fixes to all vulnerable major versions makes \emph{floating-minor} no better than \emph{pinning}.
For example, if the dependents of \pkgname{dset} use the \emph{floating-minor} constraint, i.e., $\wedge0.0.0$ or $\wedge1.0.0$ or $\wedge2.0.0$, they will always have a vulnerable version of \pkgname{dset} in their project since the fix was not backported.
Our observation that patches are not being backported aligns with the findings of Decan et al.~\cite{decan_back_2022}.
\metareview{Pinckney et al.~\cite{pinckney_large_2023} also found that some developers release security patches with minor and major version increments}.

\begin{tcolorbox}
    \textbf{Key Insights:}
    Among outdated and vulnerable dependencies, the most commonly used version constraint type is \emph{floating-minor}, with \emph{pinning} being the next most common.
\end{tcolorbox}

\subsection{\rqref{rq:switch}: How frequently do developers change the version constraint type for dependencies, and how does the change result in dependencies being outdated or vulnerable?}

This RQ explores how changes in version constraints affect whether a dependency becomes outdated or vulnerable.
First (\rqref{rq:switch}a), we want to understand how frequently developers change the \emph{version constraint type}.
Then (\rqref{rq:switch}b), we want to analyze what the trends are in such version constraint type change.
For example, we want to know if switching from a strict \emph{pinned} version to a more flexible \emph{floating-minor} constraint is common and, if so, whether this change results in outdated and vulnerable dependencies.

\noindent\textbf{\textit{\rqref{rq:switch}a.}}
We consider four types of dependency state transitions: (1)~updated to outdated; (2)~outdated to updated; (3)~remediated to vulnerable; and (4)~vulnerable to remediated.
And each transition may be triggered by one of four change scenarios:

\noindent\step{a} The dependency releases a new version. For example, the dependency releases a new version fixing a vulnerability, but the package is using a \emph{pinned} vulnerable version, so it will be considered as remediated to vulnerable transition.

\noindent\step{b} The package releases a new version but keeps the \emph{same version constraint and the same constraint type}.

\noindent\step{c} The package releases a new version with a \emph{changed version constraint} but retains the \emph{same constraint type} (e.g., remains “pinned” or “floating-minor”).

\noindent\step{d} The package releases a new version with a \emph{changed version constraint and a changed constraint type} (e.g., switches from “pinned” to “floating-minor”).

Table~\ref{table:strategy-prevalence} summarizes how often each type of change leads to the four transitions.
The most frequent cause of outdated to updated transitions is type \step{c} (over 5M cases), where package developers adjust constraints without changing the type.
However, 67,681 cases involve an actual type change (type \step{d}), where the package developers change the version constraint and also the version constraint type.

\step{a} is the least common in each of the dependency transitions.
\step{a} comprises 246 cases in updated to outdated (0.37\%), 5,645 cases in outdated to updated (0.11\%), 16 cases in remediated to vulnerable (0.64\%), and 51 cases in vulnerable to remediated (0.24\%).
\step{a} is the case when the dependency version release is auto-adopted by the package.

\step{b} plays a minimal role in regression transitions (to outdated and to vulnerable) but rarely used in upgrades (to updated and to remediated).
\step{b} comprises 1,016 cases in updated to outdated (1.52\%), 85 cases in outdated to updated ($<$0.01\%), 1,271 cases in remediated to vulnerable (50.42\%), and none in vulnerable to remediated (0.00\%).

\step{c} is the most prevalent occurrence across all transitions, particularly in upgrades.
\step{c} comprises 47,795 cases in updated to outdated (71.41\%), 5,222,276 cases in outdated to updated (98.52\%), 594 cases in remediated to vulnerable (23.55\%), and 19,549 cases in vulnerable to remediated (78.66\%).
This result indicates developers changing the version constraint but keeping the version constraint type is the most prevalent. 

\step{d} reflects deliberate restructuring (both constraint and constraint type).
\step{d} comprises 22,887 cases in updated to outdated (34.10\%), 67,681 cases in outdated to updated (1.28\%), 1,133 cases in remediated to vulnerable (44.93\%), and 3,829 cases in vulnerable to remediated (15.41\%).

\begin{table*}[ht]
\centering
\caption{Cox Time-Varying Model Results for Dependency Outdatedness and Vulnerability}
\label{tab:cox-model-comparison}
\begin{tabular}{|l||c|c|c|c||c|c|c|c|}
\hline
\multirow{2}{*}{\textbf{Covariate}} & \multicolumn{4}{c||}{\textbf{Outdatedness}} & \multicolumn{4}{c|}{\textbf{Vulnerability}} \\
\cline{2-9}
& Coef & HR (exp(coef)) & se(Coef) & p-value & Coef & HR (exp(coef)) & se(Coef) & p-value \\
\hline
floating-major        & -4.87 & 0.01 & 0.04 & $<$0.005 & --    & --    & --    & --      \\
floating-minor        & -0.82 & 0.44 & 0.01 & $<$0.005 & -0.56 & 0.57  & 0.05  & $<$0.005 \\
floating-patch        & -0.28 & 0.75 & 0.01 & $<$0.005 & 0.39  & 1.48  & 0.09  & $<$0.005 \\
fixed-ranging       & -0.48 & 0.62 & 0.01 & $<$0.005 & 0.99  & 2.69  & 0.06  & $<$0.005 \\
complex-expression    & -1.43 & 0.24 & 0.04 & $<$0.005 & 1.26  & 3.52  & 0.15  & $<$0.005 \\
at-most               & 0.00  & 1.00 & 0.02 & 0.96     & --    & --    & --    & --      \\
or-expression         & 0.66  & 1.94 & 0.03 & $<$0.005 & --    & --    & --    & --      \\
not-expression        & -3.28 & 0.04 & 0.22 & $<$0.005 & --    & --    & --    & --      \\
\hline
\end{tabular}

\end{table*}

In total, \step{a} results in 5,958 (0.11\%), \step{b} in 2,372 (0.04\%), \step{c} in 5,290,214 (98.09\%), and \step{d} in 95,530 (1.77\%) dependency state transitions.
Our results indicate that auto-adopting newer dependency version releases due to version constraint types is a rare case in changing dependency state.

\begin{tcolorbox}
    \textbf{Key Insights:}
    Updating the version constraint while retaining the same type is the most prevalent case that leads to updated or remediated dependencies.
    Retaining the version constraint type indicates that developers have a preferred version constraint type for each dependency.
\end{tcolorbox}

\noindent\textbf{\textit{\rqref{rq:switch}b.}}
We analyze type \step{d} transitions to identify the trends in constraint type change.
Packages releasing a new version with a changed constraint and a changed constraint type correspond to 1.8\% from Table~\ref{table:strategy-prevalence}.
Then, we analyze the frequency of each unique type-change pair.
Table~\ref{tab:spec-strategy-transitions} presents the top 5 frequent version constraint type changes in each transition group.

We find that developers often move from \emph{pinning} to more flexible types (e.g., \emph{floating-minor}) when updating dependencies.
Changing \emph{pinning} to \emph{floating-minor} occurred in 28,591 outdated to updated transitions and 1,154 vulnerable to remediated transitions, both of which are the highest in that transition group.
In contrast, some developers become more restrictive in changing the version constraint.
For example, changing \emph{floating-minor} to \emph{pinning} is the highest occurring pattern in transitions toward outdatedness or vulnerability.
For vulnerable to remediated transitions, removing \emph{pinning} constraints for \emph{floating-minor}, \emph{floating-major}, or \emph{fixed-ranging} appears to help exposed packages get fixed dependency versions.
By analyzing the number of occurrences for each of the cases from Table~\ref{tab:spec-strategy-transitions}, we can conclude that these transitions are not rare.

\metareview{
Under-/over-representation of each dependency type could have an impact on the way dependency constraints are present in Table~\ref{tab:spec-strategy-transitions}.
For example, \emph{floating-minor} and \emph{pinning} are frequently present in the transitions since \emph{floating-minor} and \emph{pinning} are the top-2 most frequently used version constraint types.
In addition, less frequently used version constraint types (e.g., \emph{or expression}, \emph{not expression}, \emph{complex expression}) are not present in the top-5 frequent transitions in Table~\ref{tab:spec-strategy-transitions} since these are less represented overall in version constraints.
}

\begin{tcolorbox}
    \textbf{Key Insights:}
    The top 3 version constraint type changes to remediate dependencies involved removing \emph{pinning}.
\end{tcolorbox}

\subsection{\rqref{rq:impact}: How do dependencies' version constraint type affect the dependencies' time to become vulnerable? How do dependencies' version constraint type affect the dependencies' time to become outdated?}

To interpret the models, we examine the hazard ratios: the hazard of the covariate of interest divided by the hazard of the reference group, which in our case is \emph{pinning}.
A hazard ratio $>1$ indicates increased hazard compared to the pinning category, and a hazard ratio $<1$ indicates decreased hazard compared to the pinning category.
A hazard ratio of 1 implies equal hazard between the specific version constraint and \emph{pinning}. 

\noindent\textit{\textbf{Time to Outdatedness.}}
As shown in Table~\ref{tab:cox-model-comparison}, all covariates have p-values that are $<0.005$, except for the \emph{at-most} covariate, demonstrating that the effects of all covariates (minus the at-most covariate) are statistically significant
\review{and that the observed results are unlikely due to random chance.}
The hazard ratio of 1 for \emph{at-most} indicates that dependencies with the \emph{at-most} constraint type \review{have the same hazard (instantaneous risk of becoming outdated at any given time)} as \emph{pinning}. Statistically, there is no evidence that the at-most constraint type differs from the pinning constraint type.
The \emph{or-expression} type has a hazard ratio greater than 1, indicating that dependencies with the \emph{or-expression} constraint type have about a 94\% higher likelihood of becoming outdated compared to \emph{pinning}.
The hazard ratios for the rest of the constraint types are less than 1, implying a decreased likelihood of becoming outdated compared to \emph{pinning}.
The hazard ratios of these constraint types range from 0.01 (\emph{floating-major}) to 0.75 (\emph{floating-patch}), indicating that dependencies with the floating-major constraint type have a 99\% reduction in the likelihood of becoming outdated, while those with the \emph{floating-patch} constraint type have a 25\% reduction in likelihood.

The decreasing hazard ratios for \emph{floating-patch} (0.75), \emph{floating-minor} (0.44), and \emph{floating-major} (0.01) can be explained by the relative restrictiveness compared to \emph{pinning} from Fig.~\ref{fig:versioning-spectrum}.
\emph{Floating-major} has the lowest hazard ratio, which is expected since \emph{floating-major} is the most flexible version constraint type.
Packages using \emph{floating-major} essentially keep the latest available version of that dependency.
Similarly, \emph{not-expression} has a very low hazard ratio (0.04) due to the design of \emph{not-expression}.
Packages use \emph{not-expression} to avoid a certain version of the dependency (likely because of an incompatibility issue, or breaking changes).
In doing so, the dependency resolution algorithm can pick the latest dependency version if the latest version is not the one excluded, and so \emph{not-expression} is essentially a modified form of \emph{floating-major}.
However, the hazard ratio for \emph{or-expression} is difficult to reason with, since it is not known why developers would use \emph{or-expression}, and also \emph{or-expression} is npm-specific.

\begin{tcolorbox}
    \textbf{Key Insights:}
    Version constraint type matters in becoming outdated dependencies.
    \emph{Floating-major} is the least likely and \emph{or-expression} is the most likely to become outdated.
\end{tcolorbox}

\noindent\textit{\textbf{Time to Vulnerable.}}
We removed four covariates from our vulnerability model: \emph{or-expression}, \emph{not-expression}, \emph{at-most}, and \emph{floating-major}.
First, we removed \emph{floating-major} and \emph{or-expression} as none of the dependencies with these version constraint types became vulnerable, as shown in Table~\ref{tab:occurence-percentage}.
We also removed \emph{not-expression} as only one dependency experienced vulnerability with \emph{not-expression}.
These covariates had an event rate of 0\% (or close to 0\% for the \emph{not-expression}) and would not provide useful information to the hazard estimation function.
Next, we removed \emph{at-most}, as the covariate caused convergence issues in our model.
The disproportionately high number of events occurring in the early time period (first 100 days) created a 
quasi-separation issue, which prevented a stable hazard estimation \cite{nahhas2024introduction}. 

\emph{Floating-minor}, \emph{floating-patch}, \emph{fixed-ranging}, and \emph{complex-expression} are the four covariates included in the model.
As shown in Table~\ref{tab:cox-model-comparison}, all four covariates have p-values $< 0.005$, illustrating that \review{the observed results are statistically significant and not due to random chance}.
\emph{Floating-minor} is the only covariate with a hazard ratio $< 1$.
Dependencies using the \emph{floating-minor} have a 43\% reduction in the likelihood of becoming vulnerable.
The hazard of vulnerability is 1.5, 2.7, and 3.5 times higher for \emph{floating-patch}, \emph{fixed-ranging}, and \emph{complex-expression}, respectively.

Since \emph{floating-patch} and \emph{floating-minor} are more flexible than \emph{pinning} (Fig~\ref{fig:versioning-spectrum}), we expected both to have hazard ratios $<1$.
However, \emph{floating-patch}'s 1.48 hazard ratio indicates \emph{floating-patch} is 48\% more likely to result in vulnerability.
Not backporting vulnerability fixes to all major versions (our finding from \rqref{rq:versioning}) can explain the 1.48 hazard ratio of \emph{floating-patch}.
With the same reasoning, \emph{floating-minor} should also have $>1$ hazard ratio.
In contrast to our expectation, we found that \emph{floating-minor} is 43\% less likely to result in vulnerability than \emph{pinning}.
From Table~\ref{tab:sorted-version-constraint-diversity}, we know that \emph{floating-minor} is the most used version constraint in npm and Cargo.
The 0.57 hazard ratio of \emph{floating-minor} could be explained by the disproportionate use of \emph{floating-minor} in survival dataset.
On the other hand, \emph{fixed-ranging} (2.69) and \emph{complex-expression} (3.52) have relatively higher hazard ratios for experiencing vulnerability.
Developers using \emph{fixed-ranging} may have found a suitable range that is compatible with their use, and so they may not update dependencies frequently.

\begin{tcolorbox}
    \textbf{Key Insights:}
    Version constraint type matters in becoming vulnerable dependencies.
    \emph{Floating-minor} is the least expected, and \emph{complex-expression} is the most expected to become vulnerable.
\end{tcolorbox}

\section{Discussion and Implications}
\label{sec:discussion}

\metareview{\subsection{Implications For Developers}}

\noindent\textit{\textbf{Use Automated Dependency Update.}} 
Enabling automatic dependency updates can eradicate the tedious manual task of finding vulnerable dependencies and updating to a fixed version.
The use of a package management system (e.g., npm), dependency management tools (e.g., Dependabot), or software composition analysis tools could help developers by notifying them of outdated or vulnerable dependencies.
In our survival analysis, we found that \emph{pinning} is more likely to result in outdated dependencies than any \emph{floating}. 
Thus, we recommend to avoid \emph{pinning}.
At the very least \emph{floating-patch} should be used to balance the pros and cons of getting security updates and leaving the scope of breaking changes.
\emph{Floating-minor} should be used for upstream packages whose developers are found to be generally following the SemVer policy or for \emph{feature-complete}~\cite{coelho2017modern} packages so that bug and security fixes are adopted automatically.
\emph{Floating-major} can be used for the packages for which internal breaking changes are less expected.
\emph{Fixed-ranging} or \emph{at-most} should be used when package developers are aware of a breaking change in a certain dependency's version and can specify the version constraint to avoid the versions introducing breaking changes.

\noindent\textit{\textbf{Use Lockfiles.}}
As an alternative to \emph{pinning}, lockfiles can account for specific versions (or even cryptographic hashes) of a dependency that is tested to work properly without breaking changes.
Having a lockfile can reduce the fear of breaking changes with \emph{floating}.
If breaking changes occur, developers can iterate over the dependency versions released between the version specified in the lockfile and the highest allowed version to pinpoint the version that introduced the breaking changes to facilitate effective bug and vulnerability triaging.
Additionally, lockfiles should be periodically updated to reduce the manual labor needed for the bug or vulnerability triaging phase.

\metareview{
\subsection{Implications for Researchers}
\noindent In this study, we found that \emph{floating-minor} and \emph{pinning} are the most frequently used constraint types.
From our survival analysis, we found that \emph{floating-minor} is better than \emph{pinning}, but \emph{floating-patch} is worse than \emph{pinning} for vulnerability.
The implication of these results is that floating vs pinning is a non-trivial decision for developers, and future research could explore this direction further.
We also found that updating the version constraint while keeping the version constraint type is the most prevalent change leading to updated/remediated dependencies.
Keeping the version constraint type the same could indicate that developers have a preferred constraint type for a given dependency.
Future research is needed to reveal if such a preference exists and why.
In addition, we do not know the developers' intent in changing the version constraint types, leading to updated/remediated dependencies or vice versa.
Developers may not care about outdatedness/vulnerability, and having updated/remediated dependencies could be a byproduct of entirely different decisions or design choices.
Future research could help uncover developers' intentions leading to the changes in version constraints.
}

\metareview{
\subsection{Implications for Tool Builders}
\noindent Our work highlights that developers may have a preference for version constraint type for a given dependency.
Tool builders can help developers who want to add a dependency by showing which version constraint type is mostly used for the given dependency by its dependents.
In addition, tool builders can help find bad behaviors in SemVer compliance of packages (e.g., packages that do not propagate vulnerability fixes to all major vulnerable version ranges).
A dependency's level of adherence to SemVer can help developers in selecting the appropriate version constraint type for the given dependency.
}

\metareview{\subsection{Positioning Our Work In The Context Of Existing Research}
\noindent He et al.~\cite{he_pinning_2025} conducted a study that closely aligns with ours, as both examine the \emph{pinning} vs \emph{floating} debate.
He et al.~\cite{he_pinning_2025} conducted a simulation-based analysis, whereas we conducted an empirical analysis and survival modeling.
In addition, He et al.~\cite{he_pinning_2025} used only npm packages, whereas we expanded the scope to include npm, PyPI, and Cargo packages.
Despite these differences, both studies complement each other and agree that pinning direct dependencies does not offer significant security advantages.
He et al.~\cite{he_pinning_2025} concluded that pinning direct dependencies does not provide any security benefit in an ecosystem where floating is the mainstream practice.
Similarly, our survival analysis reveals that pinning tends to result in more outdated dependencies than floating.
Both He et al.~\cite{he_pinning_2025} and our study recommend avoiding direct pinning and advocating the use of lockfiles to manage dependencies effectively.
While the methodologies differ, our study supports He et al.~\cite{he_pinning_2025}'s conclusions.
Given the rise of \emph{pinning} dependencies~\cite{he_pinning_2025}, future research could explore alternate strategies for maintaining up-to-date dependencies for large ecosystems.
}

\section{Related Work}
\label{sec:relwork}

\noindent\textbf{Dependency Versioning Practices.}
Dietrich et al.~\cite{dietrich_dependency_2019} examined 70M dependency version constraints across 17 package managers and, through a developer survey, found that developers struggle to find the right balance between using pinned dependencies and floating dependencies.
Raemaekers et al.~\cite{raemaekers_semantic_2014} conducted an empirical study on semantic versioning versus breaking changes and found that adherence to semantic versioning principles has increased only slightly over time.
Decan et al.~\cite{decan_what_2021} analyzed the SemVer compliance in package dependencies of Cargo, npm, Packagist, and Rubygems.
The authors have found that the ``wisdom of the crowd" principle is applicable when declaring dependency constraints; if other dependents of the required dependency use SemVer-compliant constraints, it is likely that the dependency is respecting the SemVer policy.
\metareview{
Pinckney et al.~\cite{pinckney_large_2023} conducted a large-scale analysis of SemVer practices in npm and found that developers' imperfect use of SemVer version constraints often blocks the propagation of security patches to downstream projects.
}
He et al.~\cite{he_pinning_2025} conducted a counterfactual analysis and simulation to study the security and maintenance impact of version constraints in the npm ecosystem.
They found that pinning direct dependencies increases the cost of maintaining vulnerable and outdated dependencies and increases the risk of exposure to malicious package updates in larger dependency graphs.
While the prior studies in this direction focus on dependency versioning practices in-the-wild, our study focuses on the impact of dependency version constraint types in becoming outdated and vulnerable dependencies.

\noindent\textbf{Outdated Dependencies.}
Kula et al.~\cite{kula_trusting_2015} analyzed the dependency adoption lag in the Maven ecosystem and found that developers are more likely to adopt the latest version of the dependency for the newly added dependencies than for updating the existing dependencies.
In another study, Kula et al.~\cite{kula_developers_2018} studied the extent to which developers update their library dependencies and found that $81.5\%$ of projects in GitHub have outdated dependencies.
Cox et al.~\cite{cox_measuring_2015} introduced the concept of ``dependency freshness'' to quantify the up-to-date state of software dependencies.
They conducted an empirical study with ``dependency freshness'' in GitHub and found that only $16.7\%$ of dependencies are up-to-date.
Derr et al.~\cite{derr_keep_2017} identified the reasons behind developers not updating their outdated dependencies through a survey and found that developers do not update dependencies due to the fear of breaking changes, lack of knowledge, and lack of motivation.
Wang et al.~\cite{wang_empirical_2020} conducted an empirical study of dependency updated intensity and delay in OSS packages and found that $50\%$ of OSS packages have at least half of their dependencies outdated.
Huang et al.~\cite{huang_characterizing_2022} analyzed usage and updates of Java packages and found that more than half of the projects have a lag of more than 500 days from the latest dependency version.
While the prior studies on outdated dependencies analyzed the trend in usage and dependency adoption lag through various metrics, no study analyzed the impact of the version constraint type used in outdated dependencies.

\noindent\textbf{Vulnerable Dependencies.}
Pashchenko et al.~\cite{pashchenko_vulnerable_2018} conducted a study to analyze vulnerable open source dependencies in SAP software and found that 81\% of the vulnerable dependencies could be fixed by updating to a newer dependency version.
In a follow-up study, Pashchenko et al.~\cite{pashchenko_vuln4real_2022} proposed a methodology, Vuln4Real, to reliably measure the extent of vulnerable dependencies and applied that to the top 500 OSS Maven dependencies from SAP.
Kumar et al.~\cite{kumar_comprehensive_2024} studied the impact of vulnerable dependencies in open-source software and found that for most programming languages, a critical vulnerability persists on average for over a year.
Mir et al.~\cite{mir_effect_2023} investigated vulnerable dependencies and their reachability on the Maven ecosystem and found that 32\% of all projects are affected by vulnerable dependencies.
Zheng et al.~\cite{zheng_closer_2023} conducted an empirical study on vulnerabilities and vulnerable packages in Rust, finding that it takes more than two years for vulnerabilities to appear on public vulnerability databases and that one-third of the vulnerabilities have no fixed commit before their disclosure.
They also found that memory safety and concurrency issues account for two-thirds of the vulnerabilities, and the vulnerable code contains statistically significantly more unsafe functions and blocks than the rest of the code.
We learned from the prior studies that have empirically evaluated vulnerabilities and vulnerable dependencies, and explored the relationship between version constraint types used by package developers and having vulnerable dependencies.

\section{Threats To Validity}
\label{sec:threats}

\noindent\textbf{External Validity.}
A limitation concerns the generalizability of our results to other ecosystems beyond those analyzed.
Each ecosystem has its own policies, practices, and developer communities that may influence dependency management in ways not fully captured in this study.
Despite this, we believe the insights derived from version constraint usage and their impact on outdated and vulnerable dependencies are broadly relevant to other ecosystems.

\noindent\textbf{Internal Validity.}
We use vulnerability data from the OSV.dev database~\cite{osv-dev}, a widely adopted source for open-source security advisories.
Nonetheless, OSV may not capture every published advisory. Vulnerabilities not present in this dataset could lead to underreporting.
Furthermore, we do not assess whether reported vulnerabilities are reachable or exploitable in practice, as discussed in the context of VEX (Vulnerability Exploitability eXchange) reports~\cite{cisa-vex}.
In our study, all vulnerabilities are treated equally, regardless of severity or CVSS (Common Vulnerability Scoring System) score.

Additionally, our analysis focuses exclusively on runtime dependencies, which packages have direct control over.
We exclude dev and optional dependencies from our measurements.
\review{
We also excluded SemVer version qualifiers (e.g., prereleases, build metadata) from our dependency resolution and our analysis, since versions with qualifiers are not fetched through typical dependency resolution (e.g, with \texttt{npm install}).
}
Finally, while some dependencies may be more critical than others, we treat all dependencies equally.
Assigning importance weights is outside the scope of this study, but it represents a meaningful direction for future work.

\section{Conclusion and Future Work}
\label{sec:conclusion}

In this study, we empirically analyzed the version constraint types used in-the-wild and also the patterns of version constraint type changes made by the developers.
We utilized survival analysis to find the relative hazard risk of \emph{pinning} to result in outdated and vulnerable dependencies in comparison to the rest of the version constraint types.
Our study shows that \emph{floating-major} is the least likely to result in outdated and \emph{floating-minor} is the most likely to result in vulnerable dependencies.
We recommend that developers avoid \emph{pinning} and use a hybrid strategy with \emph{floating} and lockfiles.
For future researchers, our study could be extended to create a used version constraint type-based metric to measure the risk of a package before developers use that package as a dependency.

\section{Acknowledgment}
This work was supported and funded by the National Science Foundation Grant No. 2207008. Any opinions expressed in this material are those of the author(s) and do not necessarily reflect the views of the National Science Foundation.

\bibliographystyle{IEEEtran}
\bibliography{websites,references}

\begin{thebibliography}{10}
\providecommand{\url}[1]{#1}
\csname url@samestyle\endcsname
\providecommand{\newblock}{\relax}
\providecommand{\bibinfo}[2]{#2}
\providecommand{\BIBentrySTDinterwordspacing}{\spaceskip=0pt\relax}
\providecommand{\BIBentryALTinterwordstretchfactor}{4}
\providecommand{\BIBentryALTinterwordspacing}{\spaceskip=\fontdimen2\font plus
\BIBentryALTinterwordstretchfactor\fontdimen3\font minus \fontdimen4\font\relax}
\providecommand{\BIBforeignlanguage}[2]{{%
\expandafter\ifx\csname l@#1\endcsname\relax
\typeout{** WARNING: IEEEtran.bst: No hyphenation pattern has been}%
\typeout{** loaded for the language `#1'. Using the pattern for}%
\typeout{** the default language instead.}%
\else
\language=\csname l@#1\endcsname
\fi
#2}}
\providecommand{\BIBdecl}{\relax}
\BIBdecl

\bibitem{dietrich_dependency_2019}
\BIBentryALTinterwordspacing
J.~Dietrich, D.~Pearce, J.~Stringer, A.~Tahir, and K.~Blincoe, ``Dependency {Versioning} in the {Wild},'' in \emph{2019 {IEEE}/{ACM} 16th {International} {Conference} on {Mining} {Software} {Repositories} ({MSR})}, May 2019, pp. 349--359, iSSN: 2574-3864. [Online]. Available: \url{https://ieeexplore.ieee.org/abstract/document/8816809?casa_token=KYtKkGU9INEAAAAA:inn0e5XiBymxMDQ-m4nXcdomVIXGorYWwbIsXyh2nipOu--OchJ3OYpq7Fon_n-1EiuLm3g03A}
\BIBentrySTDinterwordspacing

\bibitem{jafari_dependency_2022}
\BIBentryALTinterwordspacing
A.~J. Jafari, D.~E. Costa, R.~Abdalkareem, E.~Shihab, and N.~Tsantalis, ``Dependency {Smells} in {JavaScript} {Projects},'' \emph{IEEE Transactions on Software Engineering}, vol.~48, no.~10, pp. 3790--3807, Oct. 2022. [Online]. Available: \url{https://ieeexplore.ieee.org/document/9519532}
\BIBentrySTDinterwordspacing

\bibitem{zerouali_formal_2019}
\BIBentryALTinterwordspacing
A.~Zerouali, T.~Mens, J.~Gonzalez-Barahona, A.~Decan, E.~Constantinou, and G.~Robles, ``\BIBforeignlanguage{en}{A formal framework for measuring technical lag in component repositories — and its application to npm},'' \emph{\BIBforeignlanguage{en}{Journal of Software: Evolution and Process}}, vol.~31, no.~8, p. e2157, 2019, \_eprint: https://onlinelibrary.wiley.com/doi/pdf/10.1002/smr.2157. [Online]. Available: \url{https://onlinelibrary.wiley.com/doi/abs/10.1002/smr.2157}
\BIBentrySTDinterwordspacing

\bibitem{bogart_when_2015}
\BIBentryALTinterwordspacing
C.~Bogart, C.~Kästner, and J.~Herbsleb, ``When {It} {Breaks}, {It} {Breaks}: {How} {Ecosystem} {Developers} {Reason} about the {Stability} of {Dependencies},'' in \emph{2015 30th {IEEE}/{ACM} {International} {Conference} on {Automated} {Software} {Engineering} {Workshop} ({ASEW})}, Nov. 2015, pp. 86--89. [Online]. Available: \url{https://ieeexplore.ieee.org/abstract/document/7426643}
\BIBentrySTDinterwordspacing

\bibitem{bogart_when_2021}
\BIBentryALTinterwordspacing
C.~Bogart, C.~Kästner, J.~Herbsleb, and F.~Thung, ``When and {How} to {Make} {Breaking} {Changes}: {Policies} and {Practices} in 18 {Open} {Source} {Software} {Ecosystems},'' \emph{ACM Trans. Softw. Eng. Methodol.}, vol.~30, no.~4, pp. 42:1--42:56, Jul. 2021. [Online]. Available: \url{https://dl.acm.org/doi/10.1145/3447245}
\BIBentrySTDinterwordspacing

\bibitem{raemaekers_semantic_2017}
\BIBentryALTinterwordspacing
S.~Raemaekers, A.~Van~Deursen, and J.~Visser, ``\BIBforeignlanguage{en}{Semantic versioning and impact of breaking changes in the {Maven} repository},'' \emph{\BIBforeignlanguage{en}{Journal of Systems and Software}}, vol. 129, pp. 140--158, Jul. 2017. [Online]. Available: \url{https://linkinghub.elsevier.com/retrieve/pii/S0164121216300243}
\BIBentrySTDinterwordspacing

\bibitem{raemaekers_semantic_2014}
\BIBentryALTinterwordspacing
S.~Raemaekers, A.~van Deursen, and J.~Visser, ``Semantic {Versioning} versus {Breaking} {Changes}: {A} {Study} of the {Maven} {Repository},'' in \emph{2014 {IEEE} 14th {International} {Working} {Conference} on {Source} {Code} {Analysis} and {Manipulation}}, Sep. 2014, pp. 215--224. [Online]. Available: \url{https://ieeexplore.ieee.org/abstract/document/6975655}
\BIBentrySTDinterwordspacing

\bibitem{ladisa_sok_2023}
\BIBentryALTinterwordspacing
P.~Ladisa, H.~Plate, M.~Martinez, and O.~Barais, ``{SoK}: {Taxonomy} of {Attacks} on {Open}-{Source} {Software} {Supply} {Chains},'' in \emph{{IEEE} {Symposium} on {Security} and {Privacy} ({SP})}.\hskip 1em plus 0.5em minus 0.4em\relax arXiv, 2023, arXiv:2204.04008 [cs] type: article. [Online]. Available: \url{http://arxiv.org/abs/2204.04008}
\BIBentrySTDinterwordspacing

\bibitem{zahan_openssf_2023}
\BIBentryALTinterwordspacing
N.~Zahan, P.~Kanakiya, B.~Hambleton, S.~Shohan, and L.~Williams, ``\BIBforeignlanguage{en}{{OpenSSF} {Scorecard}: {On} the {Path} {Toward} {Ecosystem}-{Wide} {Automated} {Security} {Metrics}},'' \emph{\BIBforeignlanguage{en}{IEEE Security \& Privacy}}, vol.~21, no.~6, pp. 76--88, Nov. 2023. [Online]. Available: \url{https://ieeexplore.ieee.org/document/10163720/}
\BIBentrySTDinterwordspacing

\bibitem{ossf-scorecard}
``{OSSF} {Scorecard}: Build better security habits, one test at a time,'' \url{https://scorecard.dev/}, last accessed: 31-May-2025.

\bibitem{he_pinning_2025}
\BIBentryALTinterwordspacing
H.~He, B.~Vasilescu, and C.~Kästner, ``Pinning {Is} {Futile}: {You} {Need} {More} {Than} {Local} {Dependency} {Versioning} to {Defend} against {Supply} {Chain} {Attacks},'' in \emph{Proceedings of the {ACM} on {Software} {Engineering}, {Volume} 2, {Number} {FSE}, {Article} {FSE013} ({July} 2025)}, Feb. 2025, arXiv:2502.06662 [cs]. [Online]. Available: \url{http://arxiv.org/abs/2502.06662}
\BIBentrySTDinterwordspacing

\bibitem{zenodo-artifact}
``{Replication} {Package},'' \url{https://doi.org/10.5281/zenodo.15559007}, last accessed: 31-May-2025.

\bibitem{sonarcube-package}
Sonarcube, ``What is an {O}pen {S}ource {P}ackage,'' \url{https://www.sonarsource.com/learn/open-source-package/#:~:text=An%20open%20source%20package%20is,freely%20in%20their%20own%20projects.}, last accessed: 31-May-2025.

\bibitem{depsdev}
``{Open} {Source} {Insights}: Understand your dependencies,'' \url{https://deps.dev/}, last accessed: 31-May-2025.

\bibitem{shen_understanding_2024}
\BIBentryALTinterwordspacing
Y.~Shen, X.~Gao, H.~Sun, and Y.~Guo, ``\BIBforeignlanguage{en}{Understanding vulnerabilities in software supply chains},'' \emph{\BIBforeignlanguage{en}{Empirical Software Engineering}}, vol.~30, no.~1, p.~20, Nov. 2024. [Online]. Available: \url{https://doi.org/10.1007/s10664-024-10581-2}
\BIBentrySTDinterwordspacing

\bibitem{hu_empirical_2024}
\BIBentryALTinterwordspacing
J.~Hu, L.~Zhang, C.~Liu, S.~Yang, S.~Huang, and Y.~Liu, ``Empirical {Analysis} of {Vulnerabilities} {Life} {Cycle} in {Golang} {Ecosystem},'' in \emph{Proceedings of the {IEEE}/{ACM} 46th {International} {Conference} on {Software} {Engineering}}, ser. {ICSE} '24.\hskip 1em plus 0.5em minus 0.4em\relax New York, NY, USA: Association for Computing Machinery, Apr. 2024, pp. 1--13. [Online]. Available: \url{https://dl.acm.org/doi/10.1145/3597503.3639230}
\BIBentrySTDinterwordspacing

\bibitem{liu_detecting_2025}
\BIBentryALTinterwordspacing
Y.~Liu, D.~Tiwari, C.~Bogdan, and B.~Baudry, ``Detecting and removing bloated dependencies in {CommonJS} packages,'' May 2025, arXiv:2405.17939 [cs]. [Online]. Available: \url{http://arxiv.org/abs/2405.17939}
\BIBentrySTDinterwordspacing

\bibitem{alhanahnah_depsrag_2024}
\BIBentryALTinterwordspacing
M.~Alhanahnah and Y.~Boshmaf, ``{DepsRAG}: {Towards} {Agentic} {Reasoning} and {Planning} for {Software} {Dependency} {Management},'' Oct. 2024, arXiv:2405.20455 [cs]. [Online]. Available: \url{http://arxiv.org/abs/2405.20455}
\BIBentrySTDinterwordspacing

\bibitem{akhoundali_morefixes_2024}
\BIBentryALTinterwordspacing
J.~Akhoundali, S.~R. Nouri, K.~Rietveld, and O.~Gadyatskaya, ``{MoreFixes}: {A} {Large}-{Scale} {Dataset} of {CVE} {Fix} {Commits} {Mined} through {Enhanced} {Repository} {Discovery},'' in \emph{Proceedings of the 20th {International} {Conference} on {Predictive} {Models} and {Data} {Analytics} in {Software} {Engineering}}, ser. {PROMISE} 2024.\hskip 1em plus 0.5em minus 0.4em\relax New York, NY, USA: Association for Computing Machinery, Jul. 2024, pp. 42--51. [Online]. Available: \url{https://dl.acm.org/doi/10.1145/3663533.3664036}
\BIBentrySTDinterwordspacing

\bibitem{osv-dev}
``{OSV.dev} : {A} distributed vulnerability database for open source,'' \url{https://osv.dev}, last accessed: 31-May-2025.

\bibitem{github-advisory-database}
``{GitHub} {Advisory} {Database},'' \url{https://github.com/advisories}, last accessed: 31-May-2025.

\bibitem{osv-data-sources}
``{OSV:} {Current data sources},'' \url{https://google.github.io/osv.dev/data/#current-data-sources}, last accessed: 31-May-2025.

\bibitem{gu_self-admitted_2023}
\BIBentryALTinterwordspacing
H.~Gu, H.~He, and M.~Zhou, ``Self-{Admitted} {Library} {Migrations} in {Java}, {JavaScript}, and {Python} {Packaging} {Ecosystems}: {A} {Comparative} {Study},'' in \emph{2023 {IEEE} {International} {Conference} on {Software} {Analysis}, {Evolution} and {Reengineering} ({SANER})}, Mar. 2023, pp. 627--638, iSSN: 2640-7574. [Online]. Available: \url{https://ieeexplore.ieee.org/abstract/document/10123560}
\BIBentrySTDinterwordspacing

\bibitem{cao_towards_2023}
\BIBentryALTinterwordspacing
Y.~Cao, L.~Chen, W.~Ma, Y.~Li, Y.~Zhou, and L.~Wang, ``Towards {Better} {Dependency} {Management}: {A} {First} {Look} at {Dependency} {Smells} in {Python} {Projects},'' \emph{IEEE Transactions on Software Engineering}, vol.~49, no.~4, pp. 1741--1765, Apr. 2023, conference Name: IEEE Transactions on Software Engineering. [Online]. Available: \url{https://ieeexplore.ieee.org/abstract/document/9832512/authors#authors}
\BIBentrySTDinterwordspacing

\bibitem{he_migrationadvisor_2021}
\BIBentryALTinterwordspacing
H.~He, Y.~Xu, X.~Cheng, G.~Liang, and M.~Zhou, ``{MigrationAdvisor}: {Recommending} {Library} {Migrations} from {Large}-{Scale} {Open}-{Source} {Data},'' in \emph{2021 {IEEE}/{ACM} 43rd {International} {Conference} on {Software} {Engineering}: {Companion} {Proceedings} ({ICSE}-{Companion})}, May 2021, pp. 9--12, iSSN: 2574-1926. [Online]. Available: \url{https://ieeexplore.ieee.org/abstract/document/9402644}
\BIBentrySTDinterwordspacing

\bibitem{saini_investigating_2020}
\BIBentryALTinterwordspacing
M.~Saini, R.~Verma, A.~Singh, and K.~K. Chahal, ``\BIBforeignlanguage{en}{Investigating diversity and impact of the popularity metrics for ranking software packages},'' \emph{\BIBforeignlanguage{en}{Journal of Software: Evolution and Process}}, vol.~32, no.~9, p. e2265, 2020, \_eprint: https://onlinelibrary.wiley.com/doi/pdf/10.1002/smr.2265. [Online]. Available: \url{https://onlinelibrary.wiley.com/doi/abs/10.1002/smr.2265}
\BIBentrySTDinterwordspacing

\bibitem{sun_using_2023}
\BIBentryALTinterwordspacing
Y.~Sun, D.~German, and S.~Zacchiroli, ``\BIBforeignlanguage{en}{Using the uniqueness of global identifiers to determine the provenance of {Python} software source code},'' \emph{\BIBforeignlanguage{en}{Empirical Software Engineering}}, vol.~28, no.~5, pp. 1--35, Sep. 2023, company: Springer Distributor: Springer Institution: Springer Label: Springer Number: 5 Publisher: Springer US. [Online]. Available: \url{https://link.springer.com/article/10.1007/s10664-023-10317-8}
\BIBentrySTDinterwordspacing

\bibitem{miller_understanding_2025}
C.~Miller, M.~Jahanshahi, A.~Mockus, B.~Vasilescu, and C.~Kastner, ``\BIBforeignlanguage{en}{Understanding the {Response} to {Open}-{Source} {Dependency} {Abandonment} in the npm {Ecosystem}},'' in \emph{\BIBforeignlanguage{en}{International {Conference} on {Software} {Engineering}}}, 2025.

\bibitem{li_comparison_2023}
\BIBentryALTinterwordspacing
K.~Li, S.~Chen, L.~Fan, R.~Feng, H.~Liu, C.~Liu, Y.~Liu, and Y.~Chen, ``Comparison and {Evaluation} on {Static} {Application} {Security} {Testing} ({SAST}) {Tools} for {Java},'' in \emph{Proceedings of the 31st {ACM} {Joint} {European} {Software} {Engineering} {Conference} and {Symposium} on the {Foundations} of {Software} {Engineering}}, ser. {ESEC}/{FSE} 2023.\hskip 1em plus 0.5em minus 0.4em\relax New York, NY, USA: Association for Computing Machinery, Nov. 2023, pp. 921--933. [Online]. Available: \url{https://dl.acm.org/doi/10.1145/3611643.3616262}
\BIBentrySTDinterwordspacing

\bibitem{rahman_no_2025}
\BIBentryALTinterwordspacing
I.~Rahman, R.~Paramitha, N.~Zahan, S.~Magill, W.~Enck, and L.~Williams, ``No {Vulnerability} {Data}, {No} {Problem}: {Towards} {Predicting} {Mean} {Time} {To} {Remediate} {In} {Open} {Source} {Software} {Dependencies},'' Mar. 2025, arXiv:2403.17382 [cs]. [Online]. Available: \url{http://arxiv.org/abs/2403.17382}
\BIBentrySTDinterwordspacing

\bibitem{coelho2017modern}
J.~Coelho and M.~T. Valente, ``Why modern open source projects fail,'' in \emph{Proceedings of the 2017 11th Joint meeting on foundations of software engineering}, 2017, pp. 186--196.

\bibitem{pinckney_large_2023}
\BIBentryALTinterwordspacing
D.~Pinckney, F.~Cassano, A.~Guha, and J.~Bell, ``\BIBforeignlanguage{en}{A {Large} {Scale} {Analysis} of {Semantic} {Versioning} in {NPM}},'' in \emph{\BIBforeignlanguage{en}{Proceedings of the 20th {International} {Conference} on {Mining} {Software} {Repositories}}}, 2023. [Online]. Available: \url{https://www.jonbell.net/preprint/msr23-npm.pdf}
\BIBentrySTDinterwordspacing

\bibitem{survAnalysisDuration}
P.~Sentas and L.~Angelis, ``Survival analysis for the duration of software projects,'' in \emph{11th IEEE International Software Metrics Symposium (METRICS'05)}, 2005, pp. 10 pp.--5.

\bibitem{samoladas2010survival}
I.~Samoladas, L.~Angelis, and I.~Stamelos, ``Survival analysis on the duration of open source projects,'' \emph{Information and Software Technology}, vol.~52, no.~9, pp. 902--922, 2010.

\bibitem{dawson2021application}
D.~V. Dawson, D.~R. Blanchette, and B.~L. Pihlstrom, ``Application of biostatistics in dental public health,'' in \emph{Burt and Eklund's Dentistry, Dental Practice, and the Community}.\hskip 1em plus 0.5em minus 0.4em\relax Elsevier, 2021, pp. 131--153.

\bibitem{bewick2004statistics}
V.~Bewick, L.~Cheek, and J.~Ball, ``Statistics review 12: survival analysis,'' \emph{Critical care}, vol.~8, pp. 1--6, 2004.

\bibitem{steyerberg2010concepts}
E.~W. Steyerberg and T.~A. Gerds, ``Concepts in cancer survival analysis: Research questions, data, and models,'' \emph{Surgical oncology}, vol.~19, no.~2, p.~52, 2010.

\bibitem{Davidson-Pilon2019}
\BIBentryALTinterwordspacing
C.~Davidson-Pilon, ``lifelines: survival analysis in python,'' \emph{Journal of Open Source Software}, vol.~4, no.~40, p. 1317, 2019. [Online]. Available: \url{https://doi.org/10.21105/joss.01317}
\BIBentrySTDinterwordspacing

\bibitem{zhang2018time}
Z.~Zhang, J.~Reinikainen, K.~A. Adeleke, M.~E. Pieterse, and C.~G. Groothuis-Oudshoorn, ``Time-varying covariates and coefficients in cox regression models,'' \emph{Annals of translational medicine}, vol.~6, no.~7, p. 121, 2018.

\bibitem{semver-spec}
``{Semantic} {Versioning} 2.0,'' \url{https://semver.org/}, last accessed: 31-May-2025.

\bibitem{li_large-scale_2023}
\BIBentryALTinterwordspacing
W.~Li, F.~Wu, C.~Fu, and F.~Zhou, ``A {Large}-{Scale} {Empirical} {Study} on {Semantic} {Versioning} in {Golang} {Ecosystem},'' in \emph{2023 38th {IEEE}/{ACM} {International} {Conference} on {Automated} {Software} {Engineering} ({ASE})}, Sep. 2023, pp. 1604--1614, iSSN: 2643-1572. [Online]. Available: \url{https://ieeexplore.ieee.org/abstract/document/10298458?casa_token=hP2OSK48sbUAAAAA:XNFAsxNuaPffAVeomBVGw1ZQl_mG9mhYJSenVgcusJPnRO5QHk8Cpv5YSHq6nCVHToL6vnHwcA}
\BIBentrySTDinterwordspacing

\bibitem{decan_back_2022}
\BIBentryALTinterwordspacing
A.~Decan, T.~Mens, A.~Zerouali, and C.~De~Roover, ``\BIBforeignlanguage{en}{Back to the {Past} – {Analysing} {Backporting} {Practices} in {Package} {Dependency} {Networks}},'' \emph{\BIBforeignlanguage{en}{IEEE Transactions on Software Engineering}}, vol.~48, no.~10, pp. 4087--4099, Oct. 2022. [Online]. Available: \url{https://ieeexplore.ieee.org/document/9540328/}
\BIBentrySTDinterwordspacing

\bibitem{nahhas2024introduction}
R.~W. Nahhas, \emph{Introduction to Regression Methods for Public Health Using R}.\hskip 1em plus 0.5em minus 0.4em\relax CRC Press, 2024.

\bibitem{decan_what_2021}
\BIBentryALTinterwordspacing
A.~Decan and T.~Mens, ``What {Do} {Package} {Dependencies} {Tell} {Us} {About} {Semantic} {Versioning}?'' \emph{IEEE Transactions on Software Engineering}, vol.~47, no.~6, pp. 1226--1240, Jun. 2021, conference Name: IEEE Transactions on Software Engineering. [Online]. Available: \url{https://ieeexplore.ieee.org/abstract/document/8721084?casa_token=snsxTWrs_ToAAAAA:drtQbzFU1icJ0P9II1PX6eLshRSTh94VE3T1Xc6vxFOVPfP3dN3uoY4EC3mmXOpNE76BOmznnA}
\BIBentrySTDinterwordspacing

\bibitem{kula_trusting_2015}
\BIBentryALTinterwordspacing
R.~G. Kula, D.~M. German, T.~Ishio, and K.~Inoue, ``Trusting a library: {A} study of the latency to adopt the latest {Maven} release,'' in \emph{2015 {IEEE} 22nd {International} {Conference} on {Software} {Analysis}, {Evolution}, and {Reengineering} ({SANER})}, Mar. 2015, pp. 520--524, iSSN: 1534-5351. [Online]. Available: \url{https://ieeexplore.ieee.org/abstract/document/7081869}
\BIBentrySTDinterwordspacing

\bibitem{kula_developers_2018}
\BIBentryALTinterwordspacing
R.~G. Kula, D.~M. German, A.~Ouni, T.~Ishio, and K.~Inoue, ``\BIBforeignlanguage{en}{Do developers update their library dependencies?}'' \emph{\BIBforeignlanguage{en}{Empirical Software Engineering}}, vol.~23, no.~1, pp. 384--417, Feb. 2018. [Online]. Available: \url{https://doi.org/10.1007/s10664-017-9521-5}
\BIBentrySTDinterwordspacing

\bibitem{cox_measuring_2015}
\BIBentryALTinterwordspacing
J.~Cox, E.~Bouwers, M.~van Eekelen, and J.~Visser, ``Measuring {Dependency} {Freshness} in {Software} {Systems},'' in \emph{2015 {IEEE}/{ACM} 37th {IEEE} {International} {Conference} on {Software} {Engineering}}, vol.~2, May 2015, pp. 109--118, iSSN: 1558-1225. [Online]. Available: \url{https://ieeexplore.ieee.org/abstract/document/7202955}
\BIBentrySTDinterwordspacing

\bibitem{derr_keep_2017}
\BIBentryALTinterwordspacing
E.~Derr, S.~Bugiel, S.~Fahl, Y.~Acar, and M.~Backes, ``\BIBforeignlanguage{en}{Keep me {Updated}: {An} {Empirical} {Study} of {Third}-{Party} {Library} {Updatability} on {Android}},'' in \emph{\BIBforeignlanguage{en}{Proceedings of the 2017 {ACM} {SIGSAC} {Conference} on {Computer} and {Communications} {Security}}}.\hskip 1em plus 0.5em minus 0.4em\relax Dallas Texas USA: ACM, Oct. 2017, pp. 2187--2200. [Online]. Available: \url{https://dl.acm.org/doi/10.1145/3133956.3134059}
\BIBentrySTDinterwordspacing

\bibitem{wang_empirical_2020}
Y.~Wang, B.~Chen, K.~Huang, B.~Shi, C.~Xu, X.~Peng, Y.~Wu, and Y.~Liu, ``An {Empirical} {Study} of {Usages}, {Updates} and {Risks} of {Third}-{Party} {Libraries} in {Java} {Projects},'' in \emph{2020 {IEEE} {International} {Conference} on {Software} {Maintenance} and {Evolution} ({ICSME})}, Sep. 2020, pp. 35--45, iSSN: 2576-3148.

\bibitem{huang_characterizing_2022}
\BIBentryALTinterwordspacing
K.~Huang, B.~Chen, C.~Xu, Y.~Wang, B.~Shi, X.~Peng, Y.~Wu, and Y.~Liu, ``\BIBforeignlanguage{en}{Characterizing usages, updates and risks of third-party libraries in {Java} projects},'' \emph{\BIBforeignlanguage{en}{Empirical Software Engineering}}, vol.~27, no.~4, p.~90, Apr. 2022. [Online]. Available: \url{https://doi.org/10.1007/s10664-022-10131-8}
\BIBentrySTDinterwordspacing

\bibitem{pashchenko_vulnerable_2018}
\BIBentryALTinterwordspacing
I.~Pashchenko, H.~Plate, S.~E. Ponta, A.~Sabetta, and F.~Massacci, ``\BIBforeignlanguage{en}{Vulnerable open source dependencies: counting those that matter},'' in \emph{\BIBforeignlanguage{en}{Proceedings of the 12th {ACM}/{IEEE} {International} {Symposium} on {Empirical} {Software} {Engineering} and {Measurement}}}.\hskip 1em plus 0.5em minus 0.4em\relax Oulu Finland: ACM, Oct. 2018, pp. 1--10. [Online]. Available: \url{https://dl.acm.org/doi/10.1145/3239235.3268920}
\BIBentrySTDinterwordspacing

\bibitem{pashchenko_vuln4real_2022}
------, ``{Vuln4Real}: {A} {Methodology} for {Counting} {Actually} {Vulnerable} {Dependencies},'' \emph{IEEE Transactions on Software Engineering}, vol.~48, no.~5, pp. 1592--1609, May 2022, conference Name: IEEE Transactions on Software Engineering.

\bibitem{kumar_comprehensive_2024}
\BIBentryALTinterwordspacing
S.~H. B.~I. Kumar, L.~R. Sampaio, A.~Martin, A.~Brito, and C.~Fetzer, ``A {Comprehensive} {Study} on the {Impact} of {Vulnerable} {Dependencies} on {Open}-{Source} {Software},'' in \emph{2024 {IEEE} 35th {International} {Symposium} on {Software} {Reliability} {Engineering} ({ISSRE})}, Oct. 2024, pp. 96--107, iSSN: 2332-6549. [Online]. Available: \url{https://ieeexplore.ieee.org/abstract/document/10771210}
\BIBentrySTDinterwordspacing

\bibitem{mir_effect_2023}
\BIBentryALTinterwordspacing
A.~M. Mir, M.~Keshani, and S.~Proksch, ``On the {Effect} of {Transitivity} and {Granularity} on {Vulnerability} {Propagation} in the {Maven} {Ecosystem},'' in \emph{{IEEE} {International} {Conference} on {Software} {Analysis}, {Evolution} and {Reengineering} ({SANER})}.\hskip 1em plus 0.5em minus 0.4em\relax IEEE, 2023, arXiv:2301.07972 [cs]. [Online]. Available: \url{http://arxiv.org/abs/2301.07972}
\BIBentrySTDinterwordspacing

\bibitem{zheng_closer_2023}
\BIBentryALTinterwordspacing
X.~Zheng, Z.~Wan, Y.~Zhang, R.~Chang, and D.~Lo, ``A {Closer} {Look} at the {Security} {Risks} in the {Rust} {Ecosystem},'' \emph{ACM Trans. Softw. Eng. Methodol.}, vol.~33, no.~2, pp. 34:1--34:30, Dec. 2023. [Online]. Available: \url{https://dl.acm.org/doi/10.1145/3624738}
\BIBentrySTDinterwordspacing

\bibitem{cisa-vex}
CISA, ``{Vulnerability} {Exploitability} {eXchange} ({VEX}) : {Use} {Cases},'' \url{https://www.cisa.gov/sites/default/files/2023-01/VEX_Use_Cases_Aprill2022.pdf}, last accessed: 31-May-2025.

\end{thebibliography}

\end{document}